\documentclass[preprint,12pt]{elsarticle}

\usepackage{amssymb}
\usepackage{amsmath}
\usepackage{hyphenat}
\journal{Icarus}

\begin{document}

\begin{frontmatter}



\title{A Passive Probe for Subsurface Oceans and Liquid Water in Jupiter's Icy Moons}

\author[]{Andrew Romero-Wolf\corref{cor1}}
\ead{Andrew.Romero-Wolf@jpl.nasa.gov}
\cortext[cor1]{Corresponding Author. Tel +1 6263909060}
\author{Steve Vance}
\author{Frank Maiwald}
\author{Essam Heggy}
\author{Paul Ries}
\author{Kurt Liewer}
\address{Jet Propulsion Laboratory, California Institute of Technology, 4800 Oak Grove Drive, Pasadena, CA 91101, USA}

% \author{}

% \address{}

\begin{abstract}
We describe an interferometric reflectometer method for passive detection of subsurface oceans and liquid water in Jovian icy moons using Jupiter's decametric radio emission (DAM). The DAM flux density exceeds 3,000 times the galactic background in the neighborhood of the Jovian icy moons, providing a signal that could be used for passive radio sounding. An instrument located between the icy moon and Jupiter could sample the DAM emission along with its echoes reflected in the ice layer of the target moon. Cross-correlating the direct emission with the echoes would provide a measurement of the ice shell thickness along with its dielectric properties. The interferometric reflectometer provides a simple solution to sub-Jovian radio sounding of ice shells that is complementary to ice penetrating radar measurements better suited to measurements in the anti-Jovian hemisphere that shadows Jupiter's strong decametric emission. The passive nature of this technique also serves as risk reduction in case of radar 
transmitter failure. The interferometric reflectometer could operate with electrically short antennas, thus extending ice depth measurements to lower frequencies, and potentially providing a deeper view into the ice shells of Jovian moons.  
% An instrument employing this technique would require a very small amount of resources and could be added to a radar system to provide several advantages. I
% The advantage of this passive technique is that it requires a small amount of power to operate with deployable antennas that do not need to transmit high voltages providing a more flexible design that can be implemented with a low mass and stowed volume.
\end{abstract}


\end{frontmatter}

\pagebreak

% \tableofcontents

% \pagebreak
% 
% \listoffigures
% 
% % \pagebreak
% 
% \listoftables
% 
% \pagebreak
% 
% 
% 
% \clearpage
% 
% \pagebreak
% 
% \begin{table*}[h!]
% \centering
% \caption{Properties of Jupiter as a radio source.} 
% \begin{tabular}{|p{0.2\linewidth}|p{0.25\linewidth}|p{0.25\linewidth}|p{0.25\linewidth}|}
% \hline
% Location & Jupiter/galactic Flux Ratio (1-40 MHz) & Distance to Jupiter & Angular Diameter of Jupiter \\
% \hline
% Europa & 50 dB & 670,900 km & 10.8$^{\circ}$ \\
% \hline
% Ganymede & 45 dB & 1,070,400 km & 7.0$^{\circ}$ \\
% \hline
% Callisto & 35 dB & 1,882,700 km & 4.2$^{\circ}$ \\
% \hline
% \end{tabular}
% 
% \label{tbl:Jupiter_Flux} % caption for the whole figure
% \end{table*}
% 
% \pagebreak

\section{Introduction}
Subsurface oceans in Jupiter's icy moons could provide a present-day setting for extra-terrestrial life within our Solar System. Of the three Jovian icy moons, Europa is favored as having the greatest potential to sustain life, based on strong evidence for a persistent ocean directly in contact with rock. Galileo radio science measurements indicate Europa is differentiated, with a low density water-rich layer between 80 and 170 km thick (Anderson et al., 1998, Carr et al., 1998). Galileo magnetometry provides compelling evidence for a present-day ocean through the induced magnetic field (Kivelson et al., 2000, Zimmer et al., 2000). 

Estimates of the ice shell thickness of Europa are uncertain. Thermal models of the ice shell of Europa predict a thickness of $\leq$30~km (Ojakangas and Stevenson, 1989). Studies of Galileo spacecraft data have resulted in contradictory constraints for the ice shell thickness. Analysis of the Galileo magnetometer-derived oceanic conductivities, combined with radio Doppler data-derived interior models and laboratory conductivity vs concentration data, constrain the ice thickness to be  $<$15~km with a best fit value of $\sim4$~km (Hand and Chyba, 2007). Galileo imaging of pits, domes, and dark spots provide an ice shell thickness constraint of 3-10~km (Pappalardo et al., 1998). Crater analyses, also obtained from Galileo images, constrain Europa's ice thickness to $>$3~km (Turtle et al., 2001) based on the need to isostatically support central features, and to at least 19-25~km thick from the thermal state inferred from depth-size relationships (Schenk, 2002). 

The most promising technique for direct detection of subsurface oceans in Jovian icy moons is ice-penetrating radar (IPR). A dual-frequency system, such as that described by Bruzzone et al., 2011, is capable of providing high-resolution images at shallow depths ($<$5~km) and characterize the depth of the ice up to 30~km with 100~m resolution. Unambiguous observation of a subsurface ocean demands that the detection technique have as high depth sensitivity as possible. To achieve this, the use of low frequencies ($<$30~MHz) has been proposed (Bruzzone et al, 2011). The main challenges involved with IPR are surface clutter and radio absorption of the ice, which can be reduced by use of low frequencies. However, the radio loud environment of Jupiter at frequencies $<40$~MHz requires a relatively strong transmitter.

We explore a passive interferometric reflectometry technique that makes use of Jupiter's decametric (DAM) radio emission in the 1-40 MHz band. We argue that the DAM background that interferes low frequency IPR can be used as a source of ice depth sounding. This technique could be an attractive complement to a radar system because it can share the same dipole antenna and requires very low power passive components. Interferometric reflectometry could also extend the frequency band of observation to lower frequencies by operating as an electrically short dipole, further increasing the sensitivity to deep subsurface oceans. A passive measurement system could also serve as a backup to IPR in case of transmitter failure, thereby reducing the risks associated with the instrument.

Interferometric reflectometry was first applied in the Dover Heights radio astronomical observatory in the 1940's (Bolton, 1982). In that setup, an antenna placed on a cliff observed both the direct emission of a radio source and its reflection on the sea surface. The signal was autocorrelated forming a virtual two-element interferometer. The baseline formed by the sea surface reflection provided one of the first demonstrations of radio emission from discrete sources (Bolton and Stanley, 1948) along with the first identification of cosmic radio sources including Centaurus A and the Crab Nebula (Bolton 1948). It is worth mentioning that this technique was born out of limited resources, not unlike the case for deep space probes. 

The interferometric reflectometry technique is currently applied in the measurement of snow depth using GPS signals (Larson et al., 2008, Gutmann et al., 2012). The interference between the GPS signal and its subsurface reflections modulates the signal to noise ratio with a sinusoidal wave whose frequency is directly proportional the snow depth (Larson et al., 2008). The technique has been successfully demonstrated and validated by comparison with other measurements (Gutmann et al., 2012).

\begin{figure*}[h!]
\centering
\includegraphics[width=0.9\linewidth]{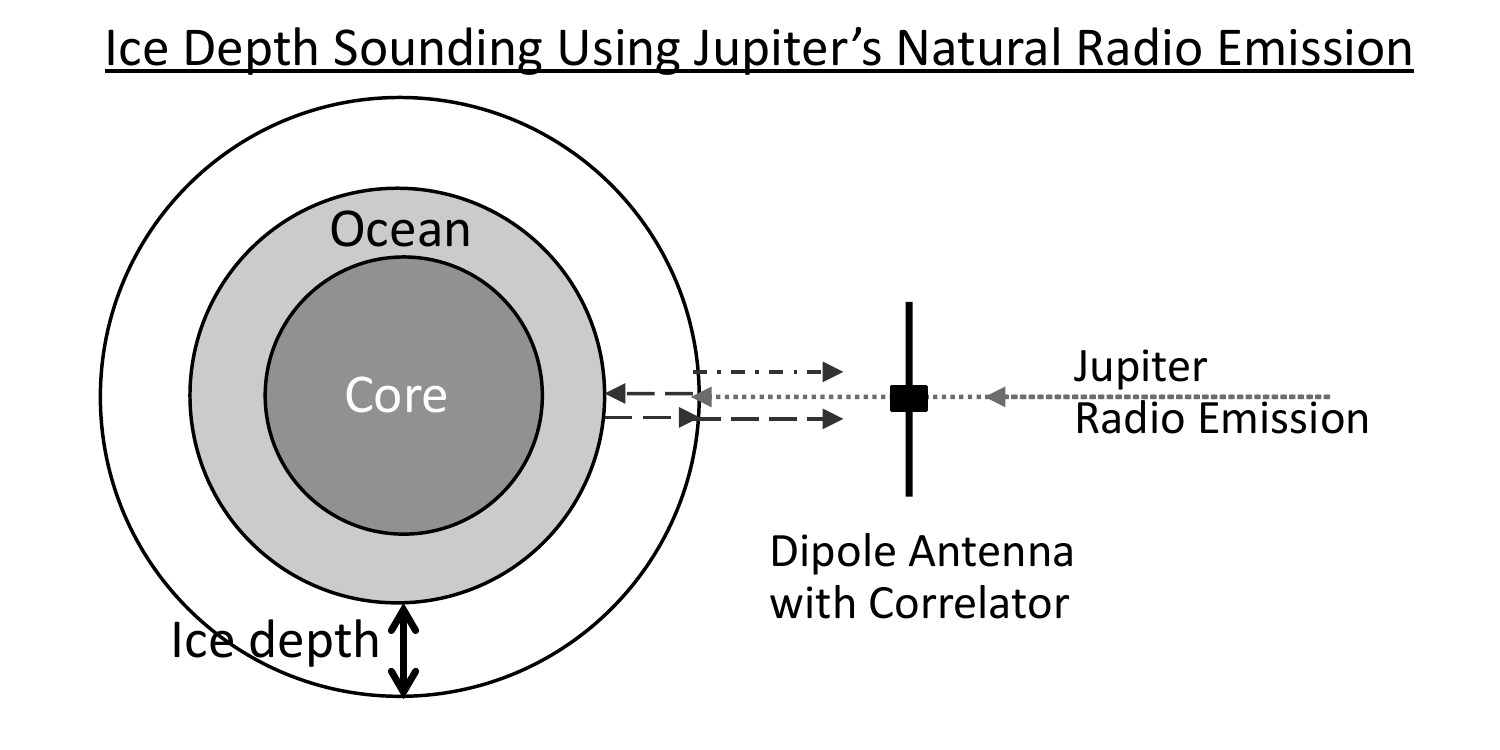}
\caption{Passive detection of subsurface oceans in icy moons using Jupiter's radio emission and its echoes. The radio emission from Jupiter (shown as an arrow with dotted line) is sampled by the dipole antenna. The radio signal is then reflected from the surface of the icy moon (arrow with dashed-dotted line) as well as the subsurface ocean (arrow with dashed line). Both echoes are detected by the dipole antenna. The delays and amplitudes of the reflected signals are extracted by correlation with the direct emission.} 
\label{fig:JANINE_concept} % caption for the whole figure
\end{figure*}

The geometry for the application of interferometric reflectometry to Jovian moon ice depth measurements is shown in Figure~\ref{fig:JANINE_concept}. Jupiter's radio emission arrives from a distance of $\gtrsim6\times10^{8}$~km to the vicinity of an icy moon. At the sub-Jovian point, where the spacecraft lies directly between Jupiter and the icy moon, an antenna receiver system records a sample of the decametric radio emission. The same emission strikes the surface of the icy moon and its echoes arrive at the spacecraft at a later time ($\sim$1~ms). The antenna beam pointed at the icy moon samples the echoed radio emission, which is cross-correlated with the direct emission to produce fringes. The cross-correlation peaks at delays corresponding to the moon surface and subsurface ice-water boundary reflection layers.  The amplitudes of the cross-correlation peaks are related to the dielectric properties of the ice. Therefore, the cross-correlation has the potential to reveal the presence of sharp boundaries 
that would be associated with a subsurface ocean or liquid water deposits in the ice shell.

In this paper, we will describe the physics of the passive interferometric reflectometer concept. In Section~2 we briefly review the properties of Jupiter's decametric radio emission. Section~3 gives a summary of the properties of Jovian moon ice shells. Section 4 describes the mathematical details of interferometric reflectometry and provides estimates for the sensitivity and resolution of the technique. Section 5 compares the expected sensitivity of interferometric reflectometry with ice penetrating radar. Section 6 summarizes our results and outlines the next steps in the development of this measurement technique.

\section{Jupiter's Decametric Radio Emission}
\subsection{Signal Strength}

Jupiter's decametric radio emissions are some of the brightest signals in the Solar system for frequencies between 1-40~MHz. The strength of the signal is due to a resonance interaction called the Cyclotron Maser Instability (Wu and Lee, 1979; Treumann 2006). The emission has a sharp cutoff at 40 MHz, which corresponds to the electron cyclotron frequency for the Jovian magnetic field lines. Radiation above 40 MHz, due to the synchrotron emission of electrons in Jupiter's magnetic field lines, is significantly weaker.

\begin{figure*}[h!]
\centering
\includegraphics[width=0.9\linewidth]{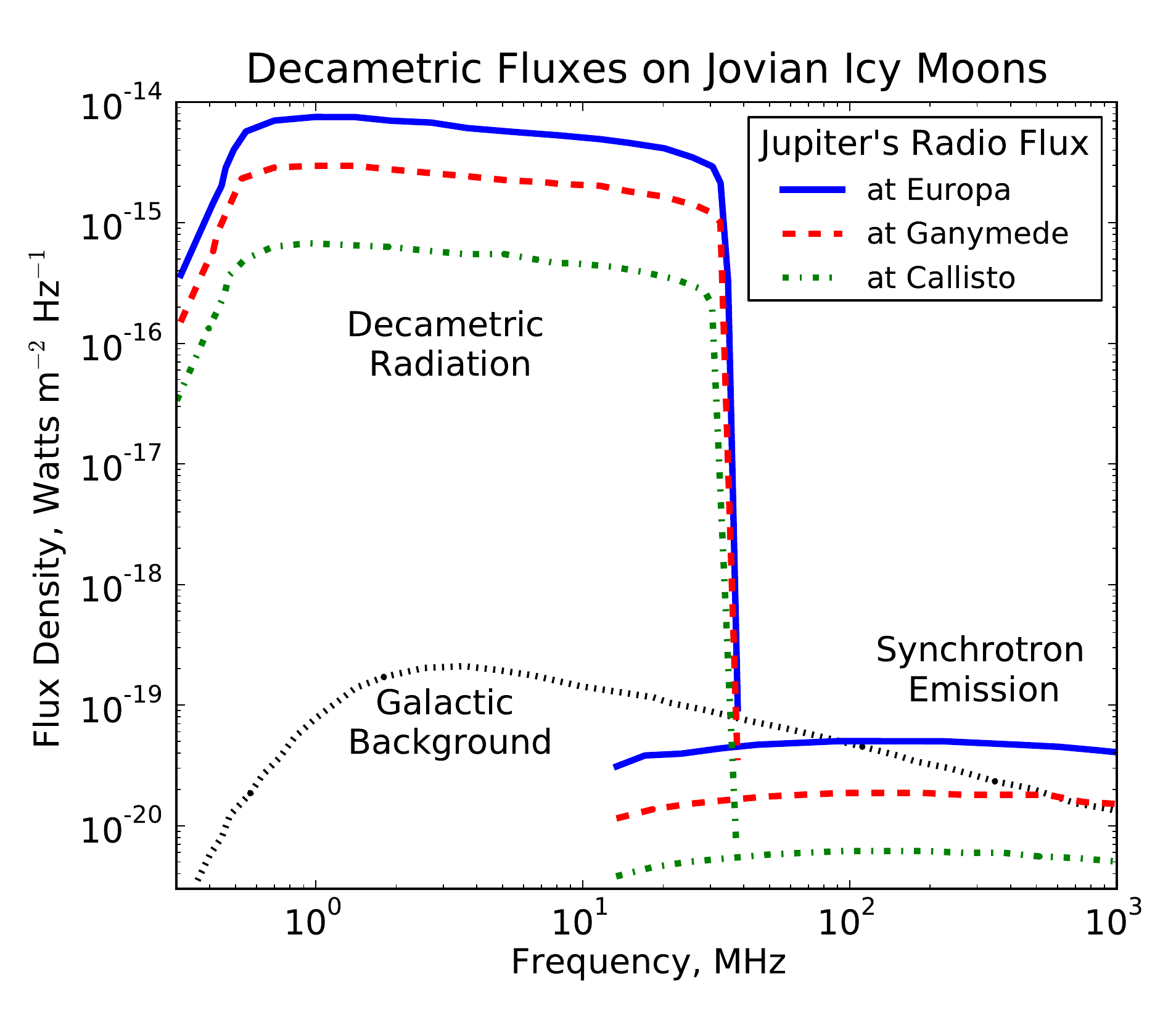}
\caption{The decametric flux density of Jupiter's radio emission in the vicinity of its icy moons far exceeds the galactic background. The curves shown are for the peak hectometric radiation ($<$3~MHz) and the decametric radiation ($3-40$~MHz) due to the Io and non-Io sources. The decametric S-bursts (not included in this Figure) can exceed the flux shown here by a few more dB. The figure is adapted from Cecconi et al., 2012.} 
\label{fig:Jupiter_Flux} % caption for the whole figure
\end{figure*}

The flux density of Jupiter's radio emission, as seen from Europa, Ganymede, and Callisto, is shown in Figure~\ref{fig:Jupiter_Flux}. Below 40 MHz, the decametric radiation from Jupiter is more than several thousands of times above the galactic background. In this frequency band, Jupiter is the most luminous object in the sky. Unlike the galactic background, which is diffuse, Jupiter's brightness distribution is confined to a small region in the sky seen by the icy moons (see Section 2.2). As will be discussed in Section 4.6, this means that the depth resolution will not be source structure limited.

\subsection{Spatio-temporal Characteristics}
% The current best constraints from Nigl, 2007 indicate that the emission region is $<$0.1 times smaller than the angular diameter of Jupiter, in one axis. The theoretical estimate of Zarka, 1996 indicates that the emission region is 0.01 times the angular diameter of Jupiter, 

There are several different sources of strong decametric emission from Jupiter, each with different characteristics.  The most reliable emission is that near the central longitude of 270$^{\circ}$.  This source (referred to as ``Non-Io-A") is not the strongest, but it occurs nearly 100\% of the time (Carr et al., 1983).  The second most active source is ``Non-Io-B", which is located at a central meridian longitude of 90$^{\circ}$-180$^{\circ}$. This source is active approximately 40\% of the time (Carr et al., 1983). 
 
The strongest source of emissions by far is a feature that occurs when Io is at 90$^{\circ}$ phase relative to the observer and the central meridian longitude of Jupiter is 90$^{\circ}$-180$^{\circ}$.  This source is referred to as ``Io-B" because it is caused by the interaction between Io and the Non-Io-B source (Thiemann, 1977).  The Non-Io-B source is active about 40\% of the time (Thiemann, 1977). An Io-A source is also active at the Non-Io-A longitude when Io is at 240$^{\circ}$ of orbital phase (Thiemann, 1977). 
 
All of the above sources vary on time scales of minutes and in amplitude by about 20 dB (Carr et al., 1983) which are referred to as long bursts (L-bursts).  However, the Io sources all produce stronger bursts than the non-Io sources and also produce short bursts (S-bursts) of duration less than 1~second that are the strongest sources of all.  These bursts make up less than 10\% of the bursts emitted from Jupiter (Carr et al., 1983). These sources are also quite small, with evidence from the very first VLBI observations showing both the S-bursts and L-bursts come from a region smaller than 400~km (Dulk 1970, Carr et al., 1970, Lynch et al., 1976).

The presence of signals with various characteristics can be used for passive radio detection. The predictability of their location and time of activity will be beneficial for planning observations. The S-bursts and L-bursts are particularly attractive due to the highly localized source and temporal structure. In this study we will use a thermal noise model for the phase of the signal. As will be discussed in Section 4, the integration time required for this passive measurement is comparable to the duration of the S-bursts and L-bursts. It is likely that the spectro-temporal characteristics of these bursts (Zarka 1996, Zarka 2004), not included in this study, will improve the estimates made with our thermal noise model, and will be the subject of a future study.
 
% \begin{table}
% \small
% \begin{tabular}{|p{0.14\linewidth}|p{0.12\linewidth}|p{0.12\linewidth}|p{0.12\linewidth}|p{0.17\linewidth}|p{0.12\linewidth}|}
% \hline
% Source & \nohyphens{Frequency Band} & \nohyphens{Active Region} & \nohyphens{Temporal Scale} & \nohyphens{Probability of Occurrence} & \nohyphens{Peak Strength} \\
% \hline
% Non-Io-A & & & & & \\
% Non-Io-B & & & & & \\
% Io-A & & & & & \\
% Io-B & & & & & \\
% \hline
% \end{tabular}
% \end{table}

\section{Properties of the Jovian Moon Ice Shells}

\subsection{Surface Properties}

Galileo and Voyager spacecraft images reveal a variety of ice surface features of Europa and Ganymede (Greeley et al., 2004). Some of these features indicate that the surface of these moons were once or may currently be geologically active (Schmidt, 2011). For a radio probe, this means that the surface clutter, due to features of size comparable to or greater than the wavelength, can be a potentially large source of signal loss. This is the case both for radar measurements and the passive interferometric reflectometer concept presented here. The diffusion effects of surface clutter drive the use of long wavelengths in the decametric range (Bruzzone et al., 2011). This plays a significant role in the detection sensitivity of oceans beneath a thick shell of ice.

Surface roughness, due to features smaller than the wavelength of the radio probe, is typically treated with a fractal model. Bruzzone et al., 2011 estimated surface roughness losses on Ganymede. The results are, however, of a speculative nature since there is not sufficient data on the surface characteristics at decametric scales. In Bruzzone et al. 2011, the sensitivity to subsurface oceans is 20~dB higher at 20~MHz compared to 50~MHz due to surface roughness and clutter effects. If the interferometric reflectometry technique can perform observations at lower frequencies (perhaps as low as $\sim$3~MHz), this could potentially provide large gains in sensitivity for subsurface ocean detection. It is worth mentioning that the use of low frequencies is highly recommended by Eluszkiewicz, 2004 based on considerations of the existence and thickness of a regolith on the surface of Europa.

We will not include surface roughness in the estimates discussed here. The losses are similar to those presented in radar studies and the reader can refer to Bruzzone et al., 2011 and Berquin et al, 2013 for more information. 
% It is worth noting that an interferometric reflectometer could operate with an electrically short dipole, thus extending its band of operation to lower frequencies, which could provide large gains in sensitivity. This will be the subject of a future study.

% \begin{equation}
% \end{equation}
\subsection{Ice Properties}
The attenuation length of the Jovian moon ice shells has not been measured and current expectations are largely model dependent. In this study we will treat the attenuation as a free parameter and assume no particular model. We treat the absorption losses of radio signal propagation $L$ as an exponential function 
\begin{equation}
L=\exp\left(-\frac{D}{\lambda_A}\right),
\label{eqn:attenuation}
\end{equation}
where $D$ is the distance of the radio wave propagates in the ice and $\lambda_A$ is a characteristic attenuation length of the ice. The parameter $\lambda_A$ corresponds to one e-folding loss in radio signal power, which is equivalent to a 4.3~dB loss\footnote{The conversion between $\lambda_A$ and loss $\alpha$ in dB/km, which is a common parameter provided in the literature, is given by  $\alpha=4.3$~dB$/\lambda_A$.}.

The ice attenuation models of Europa by Chyba et al., 1998 vary by an order of magnitude. The equivalent total two-way attenuation modeled by Chyba et al., 1998 correspond to attenuation lengths $\lambda_A\sim$0.2~km, on the pessimistic side, and $\lambda_A\sim$2.6~km on the optimistic side. Moore, 2000 quotes a range of $\alpha\sim$9-16 dB/km corresponding to attenuation lengths of $\lambda_A\sim$0.25-0.5~km. Given that radio absorption of the ice layers are largely unconstrained, we will treat the attenuation length as an unknown in our estimates and use the results of current models as indicative of the plausible values of $\lambda_A$.

The thickness of the ice shell $d$ will also be treated as an unknown in this study. As mentioned in the introduction, data constraints predict a range ice thicknesses between $\sim$3~km (Pappalardo, 1998) and $\lesssim$30~km (Ojakangas and Stevenson, 1989).  

In this study we will model the index of refraction of the ice with $n_{ice}=1.77$ (Chyba, 1998). The index of refraction is particularly relevant in the estimation of transmission and reflection coefficients. As will be discussed in Section 4, the dielectric properties of ice limit the operation of the passive concept to a region of $\pm30^{\circ}$ around the sub-Jovian point of the icy moon. It is worth mentioning that this is the region where ice penetrating radar measurements are most severely affected by Jovian decametric radio emissions. 

\subsection{Subsurface Oceans and Liquid Water}
In the nominal scenario for radio detection of subsurface oceans and liquid water, there is a highly reflective boundary with the ice, which provides a sharp transition between the ice index of refraction $n_{ice}=1.77$ (Chyba, 1998) and the ocean water index of refraction $n_{ocn}=9.3$ (Bruzzone et al., 2011). We will make our sensitivity estimates based on this assumption.

Regions of the ice containing unfrozen brine inclusions will be subject to strong reflections dependent on the dielectric properties of the brines, an area of active research for Antarctic ices (MacGregor et al. 2007).  Marion et al., 2005, suggest that salt inclusions within the ice are thermodynamically stable only in the lower 10\% of a conductively cooling ice shell of 20~km thickness. This principle holds for any thickness of or salinity, as we have confirmed using FREZCHEM (Marion et al., 2005) assuming a temperature-dependent thermal conductivity as per Vance et al. 2014. Closer to the moon’s surface, salt inclusions could be present as trapped solid precipitates. These should move downward over time through some combination of solid-state diapirism (Pappalardo and Barr 2004, Quick et al. 2013) and possibly two-phase melt transport (Kalousova et. al., 2014). The existence of near surface melts has been be inferred from surface geological features, notably by Schmidt et al., 2011.  Active chaotic 
terrains may produce transient liquids within 1~km of the surface.  At such shallow depths, liquid brines would stand out clearly in radio measurements.

% Several studies have considered the interface between the icy lithosphere and a potential subsurface ocean in Europa. Marion et al., 2005, demonstrate application of FREZCHEM software to thermodynamically stable salt inclusions within the ice. In their analysis only the lower 10\% of a conductively cooling ice shell of 20~km thickness will contain stable brine inclusions. We have reproduced their result using FREZCHEM, and find that the lower-10\% rule applies for any ice thickness and salinity. Higher up in the ice shell, salt inclusions could be present as  trapped precipitates.  The dielectric parameters and reflection coefficient of the layer has not been quantified for either type of inclusion, though some progress has been made in quantifying radio attenuation for salt and acid concentrations found in antarctic ice to temperatures as low as 193 K (MacGregor et al. 2007).

It is also possible to look for perched liquids in shallower ice. A study of the chaos terrains on the surface of Europa estimates that there is liquid water within a few km beneath the surface (Schmidt et al., 2011). At these depths, the detection of the existence of liquid brines would manifest itself as a bright reflection due to the sudden change in index of refraction.

Over geologically short time scales it might be expected that brine inclusions should flow outward gravitationally, as investigated for Europa in the context of perched brines by Sotin et al. (2002), Nimmo and Giese (2005), and in the context of convective ice diapirs by Quick et al., 2013. 

\section{Passive Measurement Concept}
This section describes the experiment model for passive ice depth sounding using Jupiter's radio emission. In this study, we treat the ice layer as a smooth dielectric sphere with a sharp transition to a reflective subsurface ocean. We include the losses due to geometric factors calculated using geometric optics for a plane wave reflecting off the surface of a sphere. We then add the effects of dielectric reflection/transmission coefficients along with absorption in the next subsections. The expected results for an instrument consisting of a dipole antenna with a correlator are given in sections 4.5 and 4.6.

\subsection{Spacecraft Position Dependence on Reflected Signal Strength}
\begin{figure*}[h!]
\centering
\includegraphics[width=0.9\linewidth]{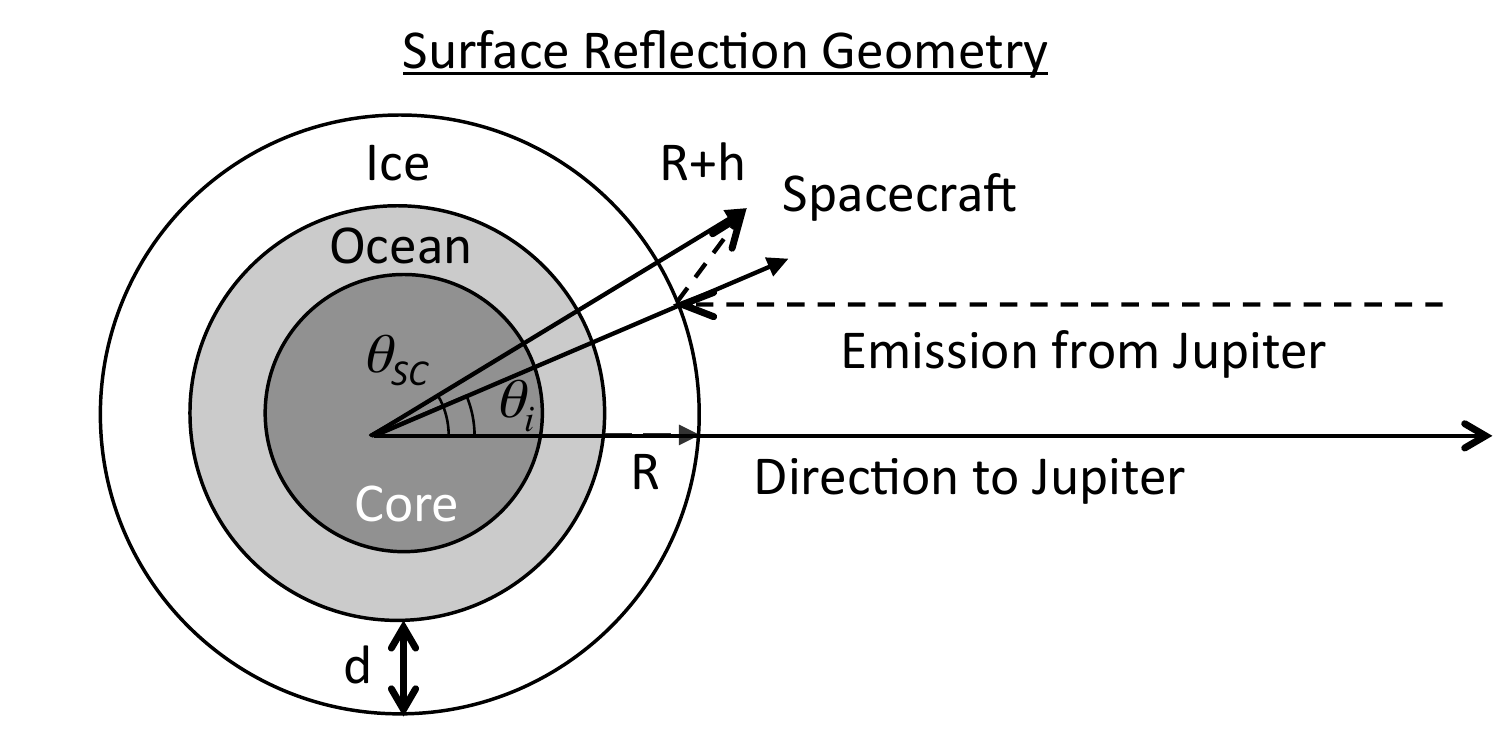}
\caption{The atmosphere-ice reflection geometry of Jovian radio emission. For a given point on the surface of the icy moon of radius $R$, the Jovian emission arrives with incident angle $\theta_i$ and is reflected to the location of a spacecraft with altitude $h$ located at angle $\theta_{SC}$.} 
\label{fig:Surface_reflection_geometry} % caption for the whole figure
\end{figure*}

In this subsection we calculate the signal losses due to the geometric factor of a plane wave reflected off a spherical surface.  We start with the geometry shown in Figure~\ref{fig:Surface_reflection_geometry} where a plane wave, coming from the direction of Jupiter, is incident on the surface of an icy moon with angle $\theta_i$. The reflection of the signal is visible from a spacecraft at altitude $h$. For this study, we are considering wavelengths in the order of $\lambda\sim$~10~-~100~meters while the radii of Jovian icy moons lie in the range of $R\sim$1560~-~2634~km. Given that $\lambda\ll R$ we use geometric optics to estimate the geometric factor losses.

Even in the geometric optics approximation, the calculation of the geometric factor is not trivial. We refer the reader to Appendix~A for a rigorous calculation of the geometric factors along with a comparison to ZEMAX (\texttt{http://www.radiantzemax.com}) simulations. We can characterize the behavior of the detector position and altitude dependence on the reflected signal strength as follows.

At the sub-Jovian point, where $\theta_i=0$, the geometric factor is given by
\begin{equation}
g(R,h)=\left(\frac{R/2}{R/2+h}\right)^2.
\label{eqn:geom_fac_subj}
\end{equation}
Note that in the limit of $h\ll R$ this expression approaches $g\sim1-4h/R$, which is a loss of order 25\% for a spacecraft at 100~km altitude over Europa. For $h\gg R$ the geometric factor approaches $g\sim h^{-2}$, which is expected from a distant source. At altitudes $h\ll R$ the geometric losses for an interferometric reflectometer are small, while for radar, the losses are always proportional to $h^{-2}$. This difference in behavior has a significant impact on the relative sensitivities of the two techniques.

\begin{figure*}[h!]
\centering
\includegraphics[width=0.7\linewidth]{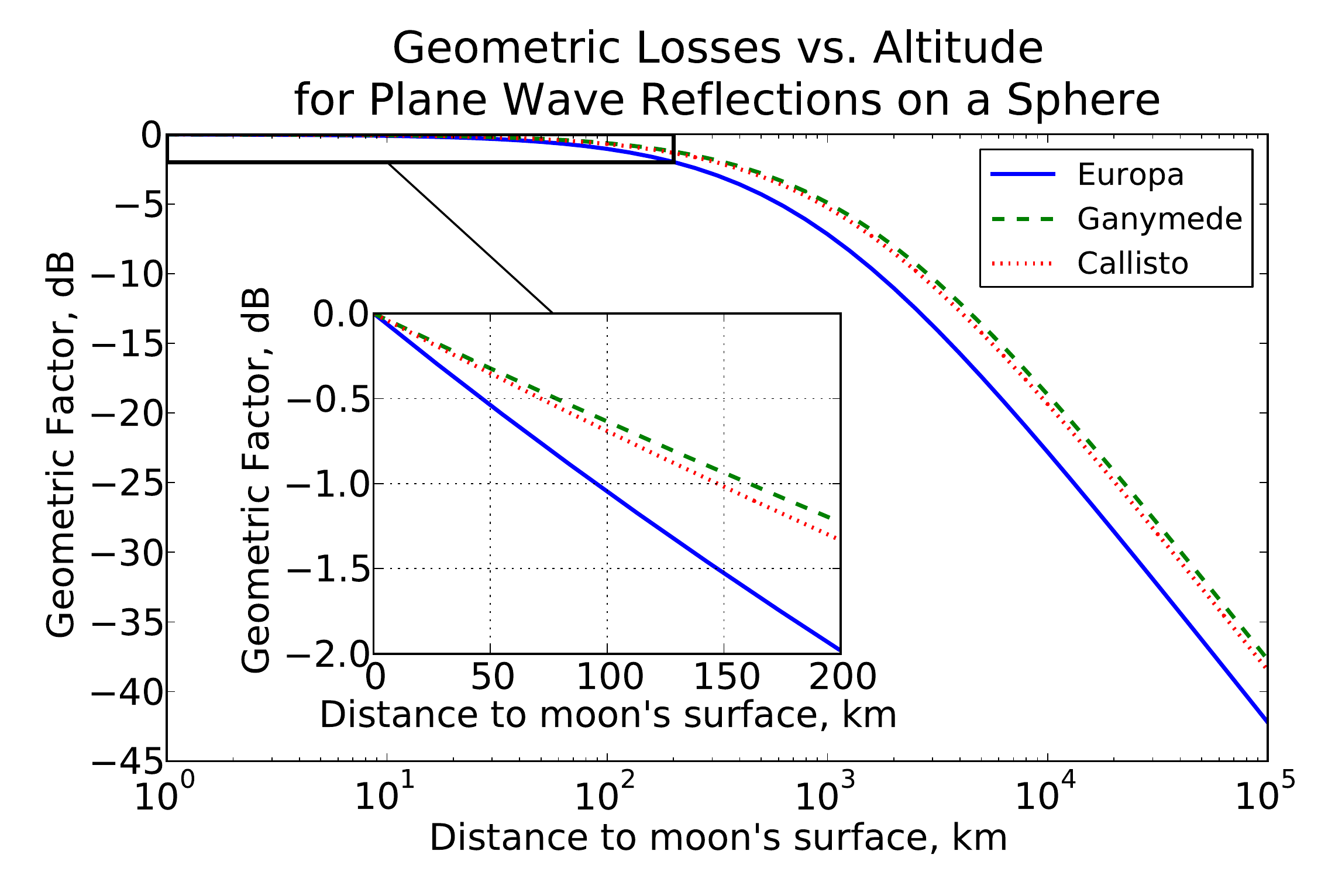}
\caption{Geometric factor as a function of spacecraft altitude between 1 and 100,000~km for Europa, Ganymede, and Callisto. The behavior of the geometric factor transitions from a linear to inverse square fall at distances comparable to the radius of the icy moon. The inset shows the altitude behavior in a linear scale for distances $<$200~km.} 
\label{fig:geom_fac_1} % caption for the whole figure
\end{figure*}

Figure~\ref{fig:geom_fac_1} plots the behavior of the losses due to the geometric factor (Equation~\ref{eqn:geom_fac_subj}) for Europa, Ganymede, and Callisto. The differences in behavior are due to the different radii of the icy moons. The behavior of the geometric factor changes at the point where the spacecraft altitude is comparable to the radius of the icy moon. For altitudes up to 200~km, the losses due to the geometric factor do not exceed 2~dB. 

\begin{figure*}[h!]
\centering
\includegraphics[width=0.7\linewidth]{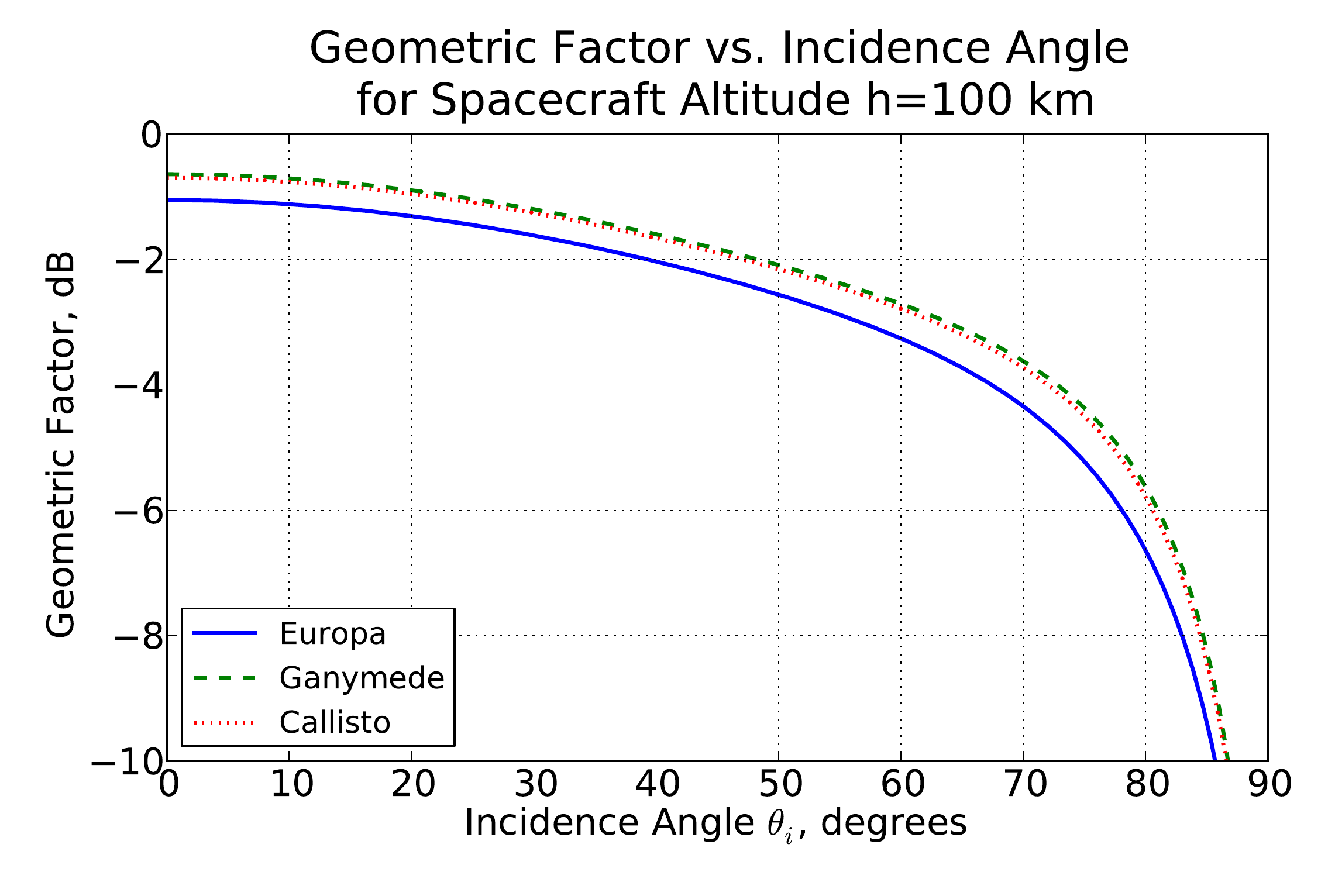}
\caption{Geometric factor as a function of incidence angle for a spacecraft altitude of 100~km on Europa, Ganymede, and Callisto. The curves differ according to the radius of each moon.} 
\label{fig:geom_fac_3} % caption for the whole figure
\end{figure*}

The incidence angle $\theta_i$ dependence on the geometric factor is derived in Appendix A (see Equation~\ref{eqn:geom_fac_calculation}). Figure~\ref{fig:geom_fac_3} plots the geometric factor as a function of incidence angle for a spacecraft altitude of 100~km over Europa, Ganymede, and Callisto. The angular dependence results in additional losses of $\lesssim$2~dB out to an incidence angle of $30^{\circ}$. At $\theta_i\sim80^{\circ}$ the losses are of 6~dB with a drastic drop at higher angles. As will be discussed in the next subsection, the refractive effects of ice allow for observations in the incidence angle range $\pm30^{\circ}$ around the sub-Jovian point. For the measurement geometries considered here, the geometric factor only contributes up to few dB of signal loss and is weakly dependent on spacecraft altitude. The incidence angle dependence of the geometric factor has been compared with a ZEMAX simulation and shows agreement to within a few percent (see Appendix A).

\subsection{Signal Intensities of Ice Surface Echoes}
As illustrated in Figure~\ref{fig:JANINE_concept}, the radio emission from Jupiter, treated as a plane wave, arrives first at the spacecraft antenna, followed by echoes from the ice surface and subsurface water. The first correlated signal of the decametric emission to arrive at the instrument will be the reflection from the atmosphere-ice boundary. Figure~\ref{fig:Surface_reflection_geometry} illustrates the reflection of plane wave arriving from Jupiter with incident angle $\theta_i$ observed by a spacecraft with altitude $h$ located at angle $\theta_{SC}$. In this figure we refer to $\theta_{SC}=0^{\circ}$ as the sub-Jovian point.

The reflection coefficients for the radio signal, arriving with incident angle $\theta_i$ from a medium with index of refraction $n_1$ towards a boundary with a medium that has index of refraction $n_2$, are given by
\begin{equation}
\rho_{\perp}(n_1, n_2, \theta_i)=\left|
\frac{
n_{1}\cos\theta_i-n_{2}\sqrt{1-\left(\frac{n_{1}}{n_{2}}\sin\theta_i\right)^2}
}
{
n_{1}\cos\theta_i+n_{2}\sqrt{1-\left(\frac{n_{1}}{n_{2}}\sin\theta_i\right)^2}
}
\right|^2
\label{eqn:refl_perp}
\end{equation}
and
\begin{equation}
\rho_{\parallel}(n_1, n_2, \theta_i)=\left|
\frac{
n_{1}\sqrt{1-\left(\frac{n_{1}}{n_{2}}\sin\theta_i\right)^2}-n_{2}\cos\theta_i
}
{
n_{1}\sqrt{1-\left(\frac{n_{1}}{n_{2}}\sin\theta_i\right)^2}+n_{2}\cos\theta_i
}
\right|^2
\label{eqn:refl_par}
\end{equation}
where $\rho_{\perp}$ is for a polarization perpendicular to the plane of incidence and $\rho_{\parallel}$ is for a polarization parallel to the plane of incidence. 

The total reflected power is given by the sum $P_{ref}=\rho_{\perp}P_{\perp}+\rho_{\parallel}P_{\parallel}$, where $P_{\perp}$ and $P_{\parallel}$ are the perpendicular and parallel polarized fractions of the radiation, respectively.   We model the surface as a layer of ice with index of refraction $n_{ice}=1.77$ and the atmosphere of the icy moon with $n_{atm}=1$. In this study we will be assuming unpolarized light so we define
\begin{equation}
\rho_{atm-ice}(\theta_i)=\frac{1}{2}\left[\rho_{\perp}(n_{atm}, n_{ice}, \theta_i) + \rho_{\parallel}(n_{atm}, n_{ice}, \theta_i)\right]
\label{eqn:refl_atm_ice}
\end{equation}

For a given Jovian flux density $S_J$ incident on the surface of the icy moon, the first reflection off the atmosphere-ice layer will depend on the reflection coefficients of the ice layer and the reflection geometric factor$\footnote{We are ignoring the effects of surface roughness for this study.}$. The flux density for a signal reflected off the atmosphere-ice boundary and arriving at the spacecraft is
\begin{equation}
S_{atm-ice}=U(R_{M},h,\theta_i)S_J
\end{equation}
where
\begin{equation}
U(R_{M},h,\theta_i)=\rho_{atm-ice}(\theta_i)g(R_{M},h,\theta_i).
\end{equation}
$\rho_{atm-ice}$ is the reflection coefficient given by Equations~\ref{eqn:refl_perp}~and~\ref{eqn:refl_par}, $g$ is a geometric factor, which is function of the radius of the icy moon $R_{M}$ and the spacecraft altitude $h$ (see Equation \ref{eqn:geom_fac_subj}). 
% For distances $h>>R_{M}$, the geometric factor behaves as $h^{-2}$. 

\begin{figure*}[h!]
\centering
\includegraphics[width=0.7\linewidth]{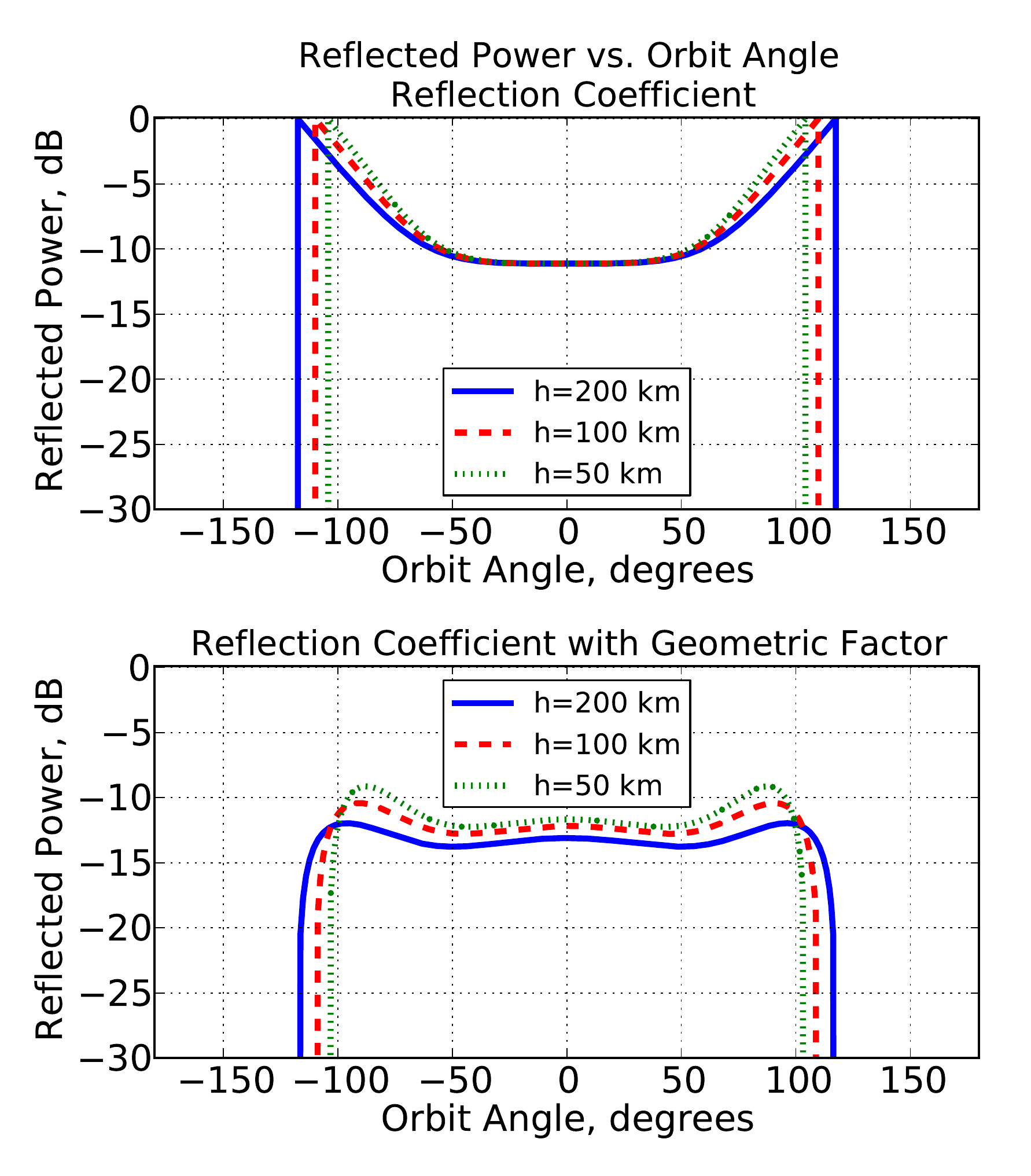}
\caption{The power reflected off a smooth sphere of ice with index of refraction $n_{ice}=1.77$. The orbit angle of $\theta_{SC}=0^{\circ}$ corresponds to the sub-Jovian point where the icy moon, the spacecraft, and Jupiter are aligned in that order (see Figure~\ref{fig:Surface_reflection_geometry}). The top panel plots the reflection coefficient as a function of spacecraft orbit angle $\theta_{SC}$ for various spacecraft altitudes.The bottom panel includes the geometric factors. The dB scale is in reference to the power incident on the surface of the icy moon. The top panel indicates what fraction of the incidence power is reflected (the rest being transmitted into the ice) while the bottom panel indicates the power detected at the spacecraft.} 
\label{fig:Surface_reflection} % caption for the whole figure
\end{figure*}

The reflection coefficient $\rho_{atm-ice}$ for the surface of a smooth icy moon as a function of spacecraft angle $\theta_{SC}$ is shown on the top panel of Figure~\ref{fig:Surface_reflection}. The three curves shown are for spacecraft altitudes of 200~km, 100~km, and 50~km. The spacecraft altitude determines the orbit angle of the horizon, shown as a sharp spikes where the curves drop abruptly where Jupiter is occulted from the spacecraft by the icy moon. For spacecraft orbit angles $\theta_{SC}<30^{\circ}$ the reflected power is nearly constant. The function $U(R_{M},h,\theta_i)$ is shown in the bottom panel of Figure~\ref{fig:Surface_reflection}. While the function $\rho_{atm-ice}$ quantifies the power reflected (where the remainder is transmitted) $U$ gives the reflected power detected at the spacecraft. The result for $\rho_{atm-ice}$, shown on the top panel of Figure~\ref{fig:Surface_reflection}, indicates that for spacecraft orbit angles $\pm30^{\circ}$ around the sub-Jovian point $\gtrsim90\%$ of 
the incident RF power is transmitted into the ice.

\subsection{Signal Intensities of Subsurface Ocean Echoes}

\begin{figure*}[h!]
\centering
\includegraphics[width=0.9\linewidth]{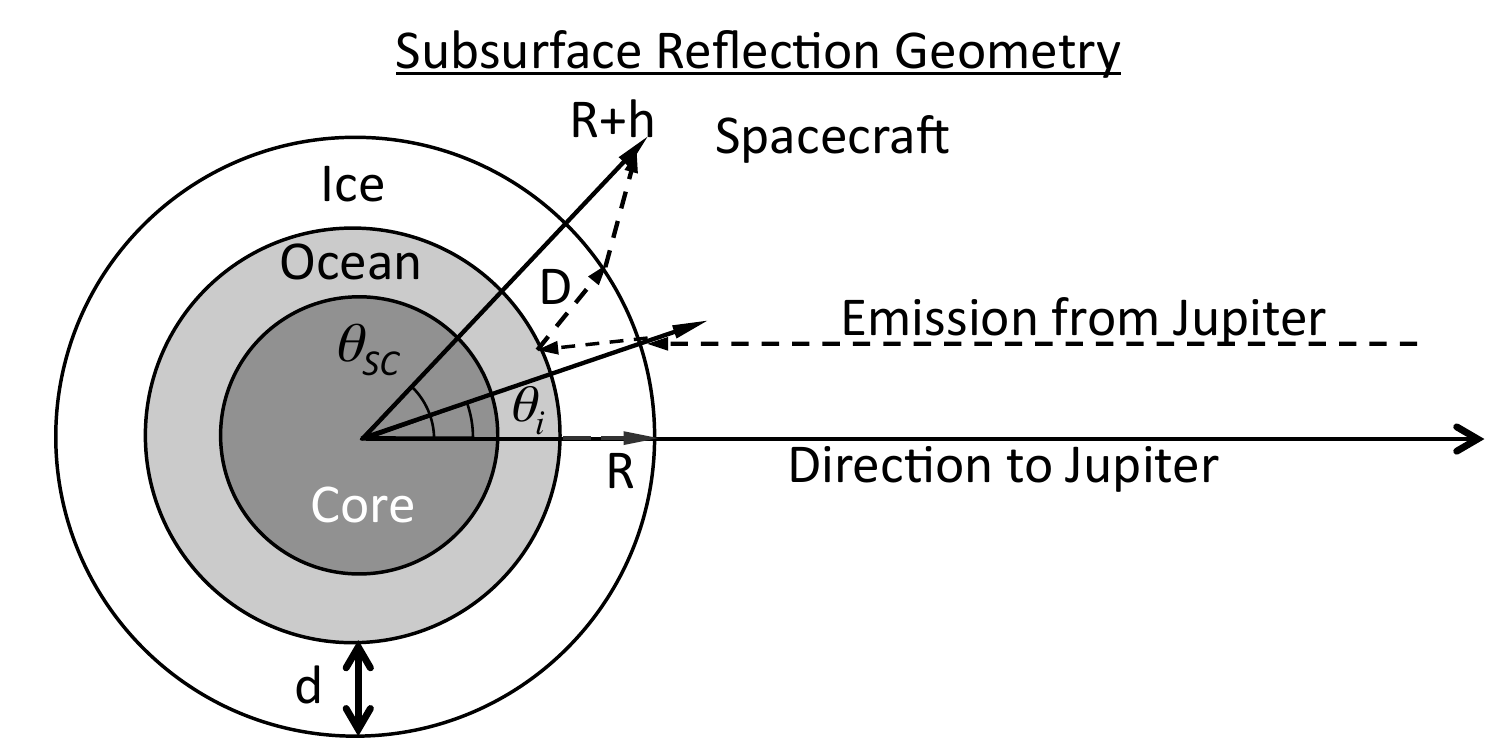}
\caption{The ice-ocean reflection geometry of Jovian radio emission. A plane wave arriving from the direction of Jupiter with incident angle $\theta_i$ is transmitted and refracted into the ice, the signal then propagates through the ice being partially absorbed until it reflects off the ice-ocean boundary. The signal propagates back out through the ice to the surface where it is transmitted and refracted out to be observed by a spacecraft. } 
\label{fig:subsurface_reflection_geometry} % caption for the whole figure
\end{figure*}

\begin{figure*}[h!]
\centering
\includegraphics[width=0.6\linewidth]{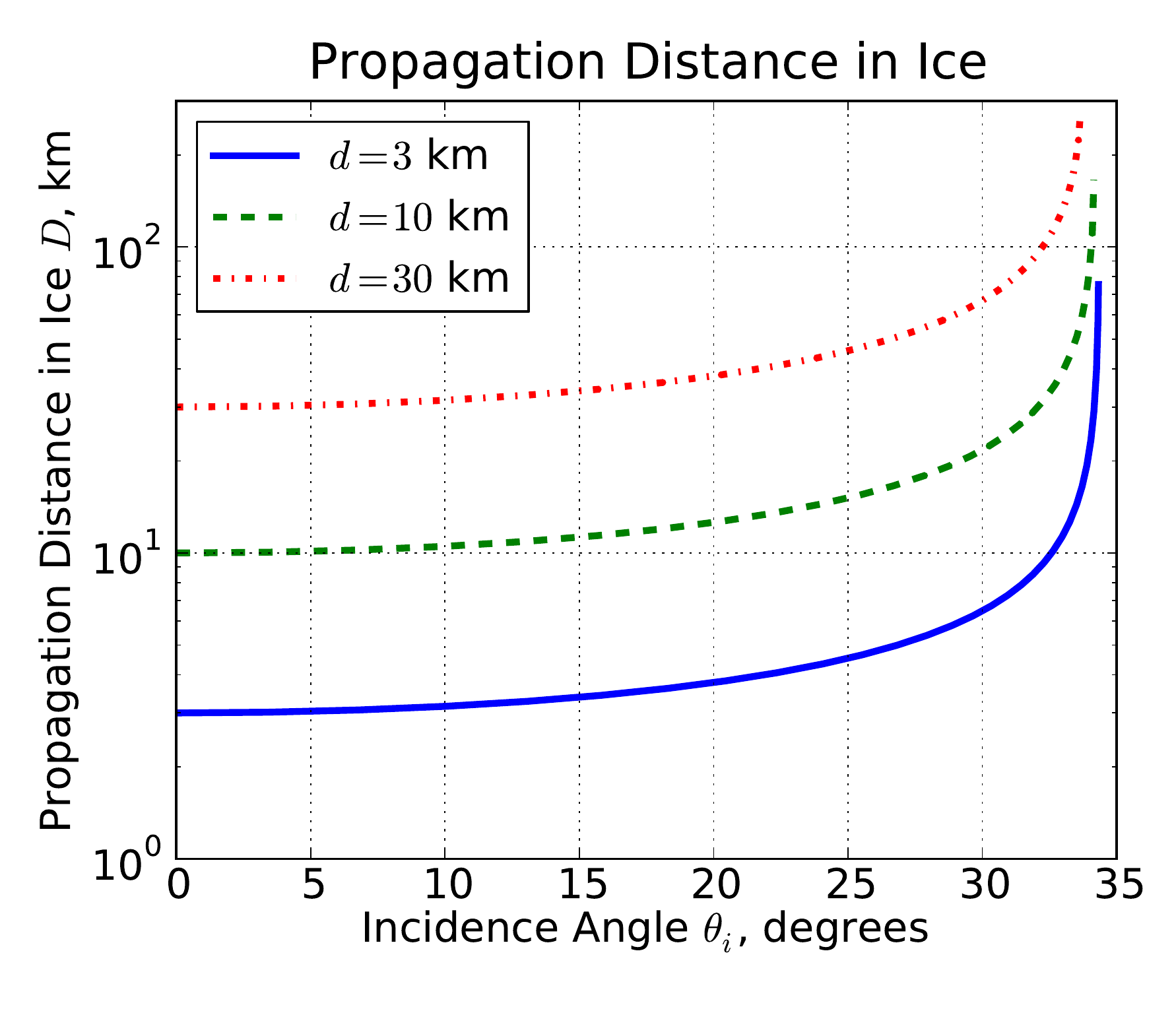}
\caption{The distance of propagation from the atmosphere-ice to the ice-ocean boundaries as a function of incidence angle $\theta_i$.  The distance $D$ is plotted for various ice depths $d$. } 
\label{fig:ice_propagation_distance} % caption for the whole figure
\end{figure*}

We estimate the amount of radio signal that undergoes a subsurface reflection and refracts out to be observed by a spacecraft. The subsurface reflection geometry is shown in Figure~\ref{fig:subsurface_reflection_geometry}. A plane wave arriving from Jupiter with incidence angle $\theta_i$ is transmitted and refracted through the atmosphere-ice boundary, where the surviving signal strength depends on the transmission coefficients $1-\rho_{atm-ice}$. The emission then propagates through the ice where it is partially absorbed according to the parameter $\lambda_A$ (see Equation~\ref{eqn:attenuation}). The distance of propagation $D$ from the surface of the ice to the subsurface layer is calculated in Appendix B (Equation~\ref{eqn:D}). The behavior of $D$ as a function of incidence angle for various ice depths $d$ is shown in Figure~\ref{fig:ice_propagation_distance}. The reflection off the ice-ocean boundary is expected to produce a smaller loss in signal due to the high index of refraction of water compared to 
ice. The reflected signal propagates a distance $D$ through the ice back to the surface, once again being partially absorbed. Finally, the emission refracts out of the ice-atmosphere boundary to be detected by the spacecraft at altitude $h$ located at angle $\theta_{SC}$.

The subsurface ocean reflection arrives at the spacecraft with flux density
\begin{equation}
S_{ice-ocn}=V(R_M, h, d, \theta_i)S_J,
\end{equation}
where
\begin{equation}
V(R_M, h, d, \theta_i)
=
\left(1-\rho_{atm-ice}\right)
\left(1-\rho_{ice-atm}\right)
\rho_{ice-ocn}
e^{-2D/\lambda_A}
g(R_{M}-d,h,\theta_r')
\end{equation}
The factor of $\left(1-\rho_{atm-ice}\right)\left(1-\rho_{ice-atm}\right)$ is due to the transmission of the radio signal entering the ice and later exiting back into free space. The reflection coefficient for the ice-ocean boundary $\rho_{ice-ocn}$  is given by Equation~\ref{eqn:refl_atm_ice} with the corresponding indices of refraction. The factor $e^{-2D/\lambda_A}$ is due to the radio absorption losses for propagation through a total propagation distance $2D$ in the ice. The geometric factor for subsurface reflections $g(R_{M}-d,h, \theta_r')$ depends on the ice depth $d$ of the icy moon and subsurface reflection angle $\theta_r'$ (see Equation~\ref{eqn:theta_r_prime} in Appendix B), but with $d\ll R_{M}$ it is a small correction to $g(R_{M},h, \theta_i)$.

\begin{figure*}[h!]
\centering
\includegraphics[width=0.7\linewidth]{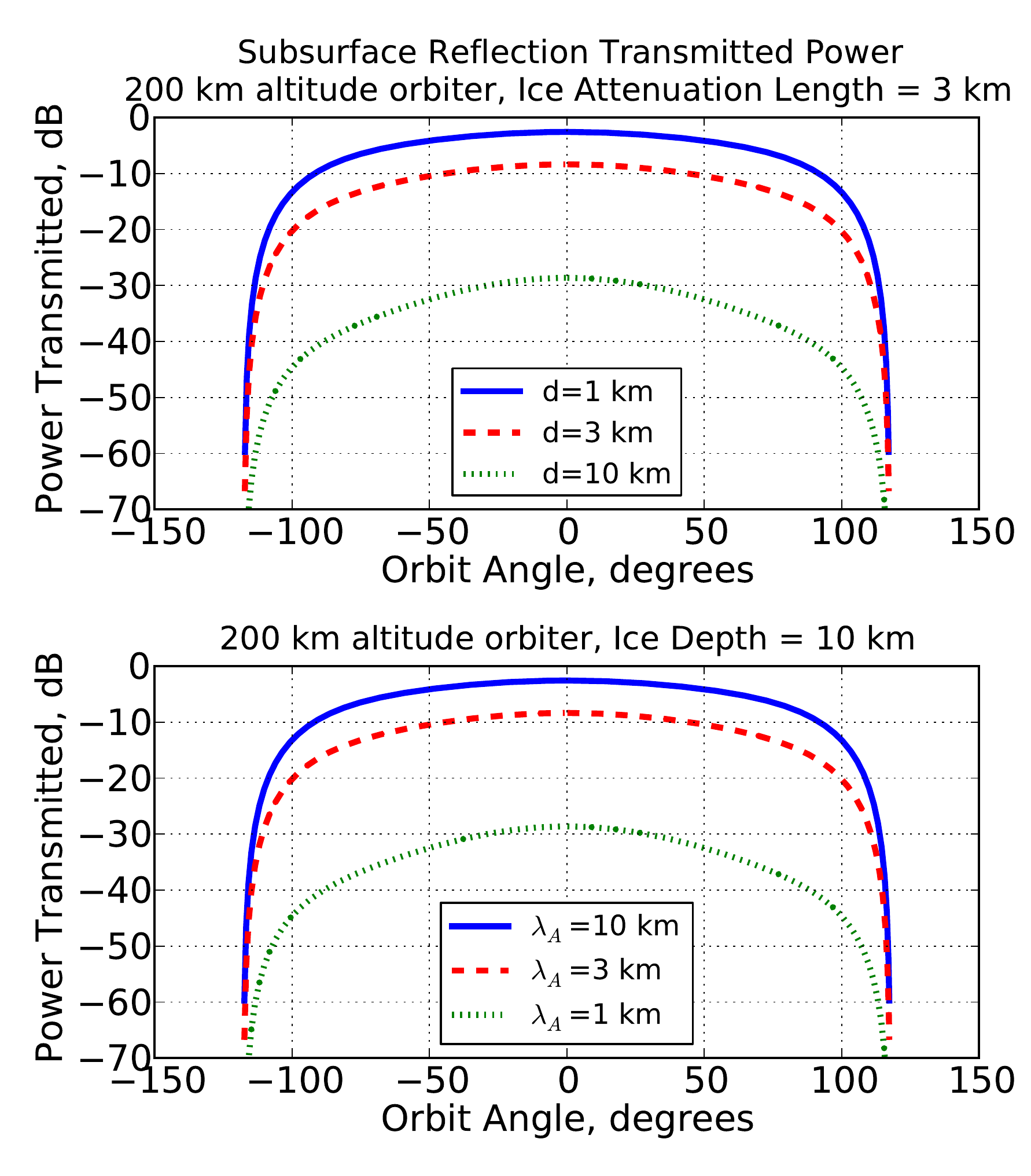}
\caption{The subsurface ice-ocean reflection power transmitted to a spacecraft assuming an unpolarized source and smooth surfaces. The curves in the top panel assume constant attenuation length and plot the absorption for various ice shell thicknesses $d$. The bottom panel assumes a constant ice depth and plots the absorption for various attenuation lengths. The dB scale is in reference to the Jovian radio emission incident on the surface of the icy moon.} 
\label{fig:subsurface_reflection} % caption for the whole figure
\end{figure*}

Figure~\ref{fig:subsurface_reflection} plots $V(R_M, h, d, \theta_i)$ vs. spacecraft orbit angle $\theta_{SC}$ (Equation~\ref{eqn:theta_SC_subsurface} in Appendix B gives $\theta_{SC}$ as a function of $R_M, h, d, \theta_i$). The curves on the top panel are for an ice attenuation length $\lambda_A=10$~km and ice depths $d$ of 1, 3, and 10~km. For deeper ice, the power loss increases orbit angle $\theta_{SC}$ due to the longer propagation lengths of the radio signal in the ice. The bottom panel of Figure~\ref{fig:subsurface_reflection} assumes an ice depth of 10~km and plots the power reflected from a subsurface ocean for attenuation lengths $\lambda_A$ of 1, 3, and 10~km. Note that different combinations of attenuation length and ice depth can give exactly the same transmitter power vs. spacecraft orbit angle making them indistinguishable by an intensity measurement alone. Fortunately, it is possible to break this degeneracy using the timing of reflections extracted by the technique discussed in the next 
subsection.

\subsection{Interferometric Reflectometry}

The cross-correlation of the radio signal arriving directly from Jupiter and its echoes off the icy moons offer the possibility of measuring the depth of a subsurface ocean. For this study, we model the signals from Jupiter as white noise. Even though it is known that Jupiter's decametric emission comes in bursts and has multiple contributors (see Section 2.2), we start with a white noise model and postpone the study of how the structure of the emission affects these measurements to a future study. The partial coherence and spectro-temporal domain structure expected from bursts is likely to improve the cross-correlation statistics compared to a white noise model.

Jupiter's radio emission with flux $S_J$ (values shown in Figure~\ref{fig:Jupiter_Flux}) contributes to the antenna temperature by an amount $T_{A,J}$ given by
\begin{equation}
T_{A,J}=\frac{A_e}{2k_B}S_{J}
\end{equation}
where $A_e$ is the effective area of the antenna and $k_{B}$ is Boltzmann's constant. 

The estimated antenna temperatures $T_{A,J}$ are shown in Table~\ref{tbl:jupiter_antenna_temperatures}, assuming a fully efficient dipole antenna with gain $G=1.7$~dBi and the fluxes shown in Figure~\ref{fig:Jupiter_Flux}. The next largest contributor to the antenna temperature is the galactic noise, which is about 4 orders of magnitude smaller (see Figure~\ref{fig:Jupiter_Flux}). A typical system noise temperature for high frequency (HF, 3-30 MHz) amplifiers is on the order of a hundred Kelvin, making it a negligible contribution. The surface of Europa has a temperature of $\sim$100~Kelvin, which is also negligible. Jupiter's radio emission is, by far, the strongest source in the vicinity of its icy moons by $\sim$40~dB above the galactic background.

\begin{table*}[h!]
\centering
\caption{The antenna temperature contribution due to Jupiter's decametric radio emission assuming a fully efficient resonant dipole antenna.}
\begin{tabular}{|c|c|c|}
\hline
Icy Moon & $T_{A,J}$ at 3~MHz in Kelvin& $T_{A,J}$ at 30~MHz  in Kelvin\\
\hline
\hline
Europa & $6.0\times10^{11}$ & $2.6\times10^9$ \\
\hline
Ganymede & $2.0\times10^{11}$ & $1.1\times10^9$ \\
\hline
Callisto & $2.0\times10^{10}$ & $2.0\times10^8$ \\
\hline
\end{tabular}
\label{tbl:jupiter_antenna_temperatures} 
\end{table*}

The time-domain radio frequency electric field amplitude at the antenna $a(t)$ can be modeled as 
\begin{equation}
a(t)=a_{J}(t)+a_{atm-ice}(t)+a_{ice-ocn}(t),
\label{eqn:amplitudes}
\end{equation}
where the Jovian electric field amplitude $a_J(t)$ is related to the direct flux density from Jupiter $S_{J}$ via
\begin{equation}
S_{J}=\frac{\left|a_{J}\right|^2}{Z_{0}}
\end{equation}
and $Z_{0}$ is the impedance of free space. The reflected amplitudes are given by
\begin{equation}
a_{atm-ice}(t)=\sqrt{U(R_{M},h,\theta_i)} \ a_{J}(t-\tau_{atm-ice})
\end{equation}
\begin{equation}
a_{ice-ocn}(t)=\sqrt{V(R_{M},h,d,\theta_i)} \ a_{J}(t-\tau_{ice-ocn})\\
\end{equation}
where $a_J(t-\tau)$ denotes the Jupiter radiation amplitude delayed by a propagation time $\tau$. The time $\tau_{atm-ice}$ denotes the delay of the direct Jovian emission and its reflection off the atmosphere-ice surface to the spacecraft. $\tau_{ice-ocn}$ denotes the delay of the direct Jovian emission and its reflection off the ice-ocean surface to the spacecraft.

\begin{figure*}[h!]
\centering
\includegraphics[width=1.0\linewidth]{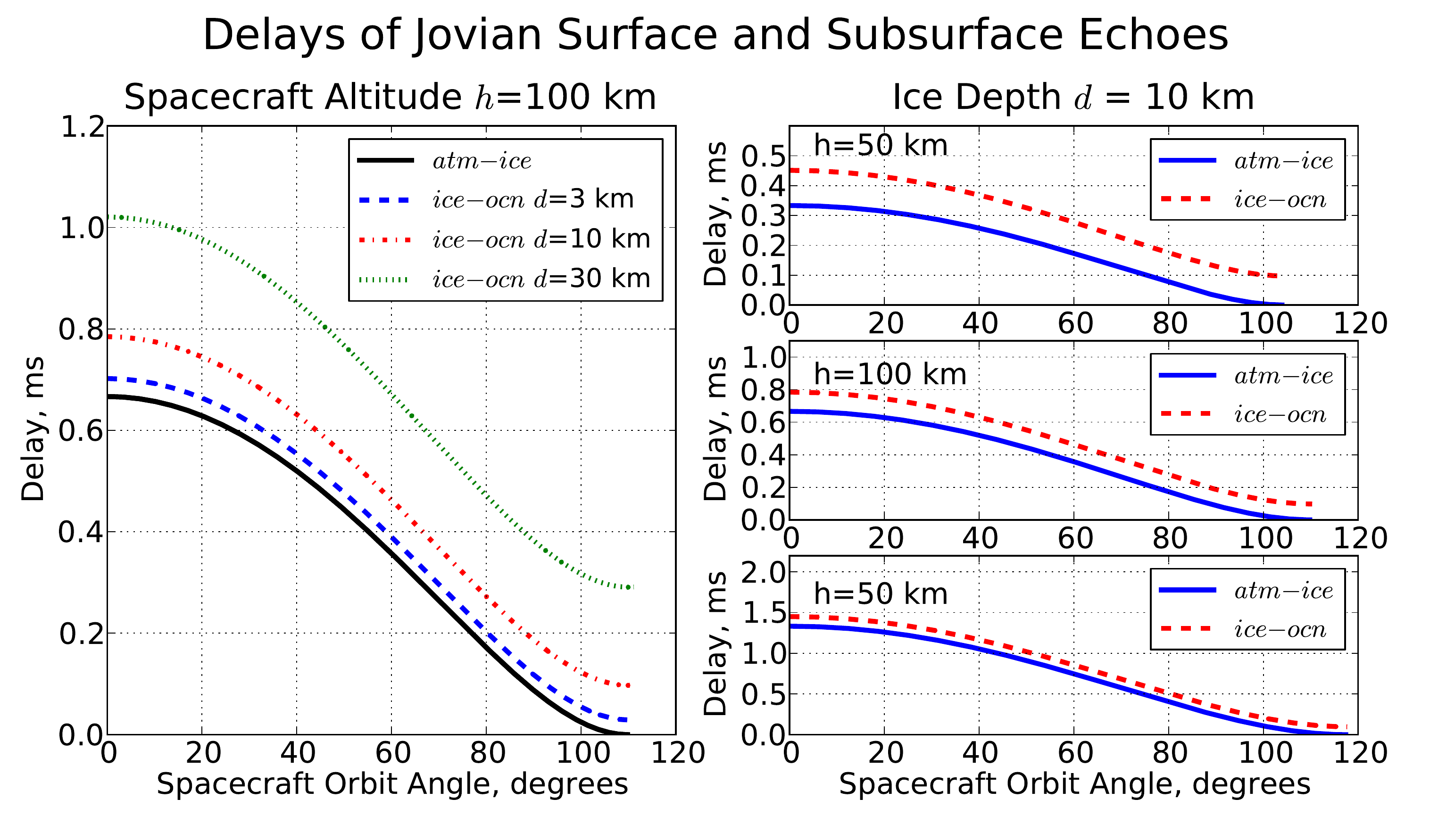}
\caption{
The geometric delay between the direct observation of Jovian emission and its surface reflection $\tau_{atm-ice}$ and subsurface reflection $\tau_{ice-ocn}$. Left: the delays are plotted for various ice depths $d$ and spacecraft orbit altitude of $h$=100~km as a function of spacecraft orbit angle $\theta_{SC}$. The difference in delays $\tau_{atm-ice}$ and $\tau_{ice-ocn}$ is directly related to the ice depth parameter. The variation in delay difference with spacecraft angle is weak ($\sim$20\%) due to the relatively high index of refraction of ice, which tends to refract light rays inwards. Right: the delays are plotted against spacecraft orbit angle for a fixed ice depth $d$=10~km for various spacecraft altitudes $h$. The difference in delays $\tau_{atm-ice} - \tau_{ice-ocn}$ is not altered significantly as a function of spacecraft altitude $h$ and relates directly to the ice depth.
} 
\label{fig:delays} % caption for the whole figure
\end{figure*}

The delays $\tau_{atm-ice}$ and $\tau_{ice-ocn}$ are plotted in Figure~\ref{fig:delays}. For a spacecraft altitude of 100~km, the delays are in the order of 1~ms. The difference in delay between $\tau_{atm-ice}$ and $\tau_{ice-ocn}$ is between 33~$\mu$s, for an ice depth of 3~km and 330~$\mu$s for an ice depth of 30~km. The difference in delays varies only at the level of 20\% between nadir and horizon spacecraft orbit angles. This is due to the relatively high index of refraction of ice. For details on the calculation of these delays see Appendix B.

The depth and surface properties can be extracted from the autocorrelation function taken over a period $T$
\begin{equation}
A(\tau)=\frac{2}{T}\int_{-T/2}^{T/2}dt \ a(t) a^{*}(t-\tau).
\end{equation}
Note that $A(0)=\left<S\right>Z_{0}$ where $\left<S\right>$ is the average total flux density on the antenna.
Using the model electric field amplitudes in Equation~\ref{eqn:amplitudes}, the autocorrelation function becomes
\begin{equation}
\begin{split}
A(\tau)  &= A_J(\tau) + A_{atm-ice}(\tau) + A_{ice-ocn}(\tau) \\ 
& + 2\left[C_{J,atm-ice}(\tau) + C_{J,ice-ocn}(\tau) + C_{atm-ice,ice-ocn}(\tau) \right]\\
\end{split}
\label{eqn:autocorrelation}
\end{equation}
where the first three terms are (1) the autocorrelation of the directly observed Jovian emission, (2) the autocorrelation of the Jovian emission reflected off the atmosphere-ice boundary, and (3) the autocorrelation of the Jovian emission reflected on the ice-ocean boundary, respectively. The last three terms of Equation~\ref{eqn:autocorrelation} are: (4) the cross-correlation of the direct Jovian emission and its reflection off the atmosphere-ice boundary, (5) the cross-correlation of the direct Jovian emission and its reflection off the ice-ocean boundary, and (6) the cross-correlation of the reflection off the atmosphere-ice boundary and reflection off the ice-ocean boundary. The cross-correlation $C_{i,j}(\tau)$ between two electric field amplitudes $a_{i}(t)$ and $a_{j}(t)$, over a period $T$, is here defined as
\begin{equation}
C_{i,j}(\tau)=\frac{2}{T}\int_{-T/2}^{T/2}dt \ a_i(t) a_j^{*}(t-\tau).
\end{equation}

The strength and timing of each of the terms in Equation \ref{eqn:autocorrelation} is listed below.
\begin{equation}
\begin{split}
A_J(\tau)  &= \langle S_J \rangle Z_0  \ \delta(\tau) \\
A_{atm-ice}(\tau) &= U(R_{M},h,\theta_i)\langle S_J \rangle Z_0 \ \delta(\tau) \\
A_{ice-ocn}(\tau) &= V(R_{M},h,d,\theta_i) \langle S_J \rangle Z_0 \ \delta(\tau) \\
\end{split}
\label{eqn:autocorrelation_terms}
\end{equation}

\begin{equation}
C_{J, atm-ice}(\tau)  = \sqrt{U(R_{M},h,\theta_i)}  \  \langle S_J \rangle Z_0 \ \delta(\tau-\tau_{atm-ice}) 
\label{eqn:cc_atm_ice}
\end{equation}
\begin{equation}
C_{J, ice-ocn}(\tau) = \sqrt{V(R_{M},h,d,\theta_i)} \ \langle S_J \rangle Z_0 \ \delta(\tau-\tau_{ice-ocn}) 
\label{eqn:cc_ice_ocn}
\end{equation}
\begin{equation}
C_{atm-ice,ice-ocn}(\tau) =  \sqrt{U(R_{M},h,\theta_i)V(R_{M},h,d,\theta_i)} \  \langle S_J \rangle Z_0 \ \delta(\tau-\tau_{ice-ocn}+\tau_{atm-ice})
\label{eqn:cc_atm_ice_ice_ocn}
\end{equation}
where $\langle S_J\rangle$ is the average power of the signal arriving directly from Jupiter and $\delta(\tau)$ is the Dirac delta function. The magnitude of the autocorrelation terms in Equation~\ref{eqn:autocorrelation_terms} give the total power of the signal including the direct emission from Jupiter and its echoes from the icy moon. Each contribution sums as a peak $\delta(\tau)$ at zero delay. The cross-correlations in Equations~\ref{eqn:cc_atm_ice}~and~\ref{eqn:cc_ice_ocn} are the interference of the direct emission with the surface and subsurface reflections, respectively. These peaks are found at the delays corresponding to the light propagation times between the spacecraft and the reflecting surfaces and arrive at separate times from each other (see Figure~\ref{fig:delays}).  Equation~\ref{eqn:cc_atm_ice_ice_ocn} is the cross-correlation due to the interference of the ice-ocean echo and the atmosphere-ice echo. It is weaker than the cross-correlation in Equations~\ref{eqn:cc_atm_ice}~and~\ref{eqn:
cc_ice_ocn} but could provide an important signature in the measurement as a distinct mark of multiple reflection layers.

Equation~\ref{eqn:autocorrelation} predicts that the autocorrelation function of a dipole lying between an icy moon and Jupiter, with an omni-directional beam pattern capable of observing both bodies simultaneously, results in interferometric peaks whose amplitude and delay are directly related to the reflective properties and locations of the surface and subsurface layers.

% Another possibility is to use a system of two antennas with more directivity that a dipole. For example, a log-periodic dipole array (LPDA) can achieve directivities of up to 9~dBi with front-to-back ratios greater than 20~dB. If we were to use two LPDA antennas, one pointed at Jupiter and the other pointed at the icy moon, this would reduce the interference from the direct emission of Jupiter on its echoes by 20~dB. In such a case, Equation~\ref{eqn:autocorrelation} would be a cross-correlation between the two antennas and the first term would be reduced by 40~dB. This is clearly a more complex system but it could provide a significantly cleaner measurement. The effect on the sensitivity of the measurement is discussed in the next subsection.

\subsection{Ocean Depth Sensitivity}

In this section we estimate the statistical limitations in the identification of  reflection peaks in the autocorrelation function (Equation~\ref{eqn:autocorrelation}). The radiometer equation states that the minimum detectable antenna temperature, at the $\sim68\%$ confidence level, is given by
\begin{equation}
\delta T = \frac{T_{sys}}{\sqrt{\Delta f \Delta t}}
\end{equation}
where $\Delta f$ is the bandwidth and $\Delta t$ is the integration time of the measurement. In general, $T_{sys}$ is a combination of the noise temperature of the electronics and undesired contributions to the antenna temperature from background sources. We have already discussed that the contributions due to system noise, the galactic background, and the surface temperature of Jovian icy moons are negligible compared to the antenna temperature contribution from Jupiter $T_{A,J}$. In this application, $T_{A,J}$ itself is the limiting background since it is continuously emitting random signals while the delayed echoed signals of interest arrive at the system. Therefore, as far as the autocorrelation function is concerned, we set $T_{sys} = T_{A,J}$ as it is the dominant contribution to the system temperature. 

It is worth noting that we are treating the decametric signal from Jupiter as white noise, which means that the autocorrelation will have a statistically random distribution of accidental correlation values, independent of frequency. This need not be the case. In fact it is known that the strongest component of the decametric emission is a non-thermal process (Treumann, 2006). Any partial coherence in the signals would likely result in an improved correlation. Characterization of the Jovian decametric burst behavior with regard to its autocorrelation statistics will be deferred to a future study.  Here we focus on estimating the sensitivity to subsurface ocean reflections with a white noise model, which is likely a more pessimistic scenario.

The term in the autocorrelation function (Equation~\ref{eqn:autocorrelation}) of interest for subsurface ocean sounding is the cross-correlation strength of the ice-ocean interface reflection with the direct Jovian emission $C_{J, ice-ocn}(\tau)$ given in Equation~\ref{eqn:cc_ice_ocn}. The equivalent temperature due to this term is 
\begin{equation}
k_{B}T_{ice-ocn}=\frac{C_{J, ice-ocn}A_{e}}{2Z_0},
\end{equation}
where $A_e$ is the effective area of the antenna. If we require a signal strength of $T_{ice-ocn}>N\delta T$ for a measurement that is $N$ times greater than background with $T_{sys}=T_{A,J}$ then
\begin{equation}
\sqrt{\Delta f \Delta t} > 
\frac{N}{\sqrt{V(R_M, h, d, \theta_i)}}
\end{equation}
The integration time needed is therefore 
\begin{equation}
\Delta t > \frac{N^2}{\Delta f \ V(R_M, h, d, \theta_i)}
\label{eqn:integration_time}
\end{equation}

Let us estimate the sensitivity to subsurface oceans for the case where the incoming plane wave, the spacecraft, and the reflection point all lie in the same axis along the sub-Jovian point. With this geometry the incident angle is $\theta_i=0^{\circ}$. The geometric factor is given by Equation~\ref{eqn:geom_fac_subj}. The reflection coefficient (Equations~\ref{eqn:refl_perp}~and~\ref{eqn:refl_par}) is given by $\rho_{\perp,\parallel}(n_1,n_2)=(n_1-n_2)^2/(n_1+n_2)^2$. In the sub-Jovian point geometry the distance of propagation in ice $D$ equals the ice depth $d$. Substituting these functions into Equation~\ref{eqn:integration_time} we have
\begin{equation}
\Delta t > 
\frac{N^2e^{2d/\lambda_A}}{\Delta f}
\left(\frac{\left(n_{ice}+n_{atm}\right)^2}{4n_{ice}n_{atm}}\right)^2
\left(\frac{n_{ocn}+n_{ice}}{n_{ocn}-n_{ice}}\right)^2
\left(\frac{R_{M}/2+h}{R_{M}/2}\right)^2
\label{eqn:nadir_integration_time}
\end{equation}

As an example, take the case of a 30~MHz dipole with 10\% bandwidth giving $\Delta f= 3$~MHz. At the sub-Jovian point, we have $\rho_{ice-ocn}=0.48$, $\rho_{atm-ice}=0.07$. For an ice depth of 10~km and an attenuation length of 3~km and a desired statistical significance of $N=5$ we have $\Delta t = 130$~ms. 

The detectability threshold of a subsurface ocean can be expressed as a ratio of the two unknown parameters $d/\lambda_A$
\begin{equation}
\frac{d}{\lambda_A} < 
\frac{1}{2}
\log\left[
\frac{\Delta t \Delta f}{N^2}
\left(\frac{4n_{ice}n_{atm}}{\left(n_{ice}+n_{atm}\right)^2}\right)^2
\left(\frac{n_{ocn}-n_{ice}}{n_{ocn}+n_{ice}}\right)^2
\left(\frac{R_{M}/2}{R_{M}/2+h}\right)^2
\right],
\label{eqn:sensitivity}
\end{equation}
where $\log$ is the natural logarithm. % where $\Delta t_{max}$ is the maximum integration time allowed by the setup of the passive interferometric reflectometer.

The limit on the integration time $\Delta t$ is due to the motion of the spacecraft over the icy moon. For a body like Europa, the orbital speed is $v\approx1.4$~km/s. As discussed in Bruzzone et al., 2011, the first Fresnel zone on the surface of the icy moon, for a spacecraft of altitude $h$ is given by
\begin{equation}
F=\sqrt{2\lambda h}
\end{equation}
For $\lambda=10$~m and $h=100$~km, we have $F=$1~km. For $\lambda=100$~m we have F=3.1~km. If we require that the spacecraft take a measurement so that the field of view is within the first Fresnel zone then the maximum integration time is
\begin{equation}
\Delta t_{max}\lesssim\frac{\sqrt{2 \lambda h}}{v}
\end{equation}
For a spacecraft altitude $h=100$~km observing at $\lambda=10$~m we have $\Delta t_{max}\lesssim 0.7$~seconds and at $\lambda=100$~m we have $\Delta t_{max}\lesssim 2.3$~seconds. 
% Note that the use of low frequencies provides a factor of $\sim$3 increase in sensitivity due to additional integration time. 

\begin{figure*}[h!]
\centering
\includegraphics[width=0.9\linewidth]{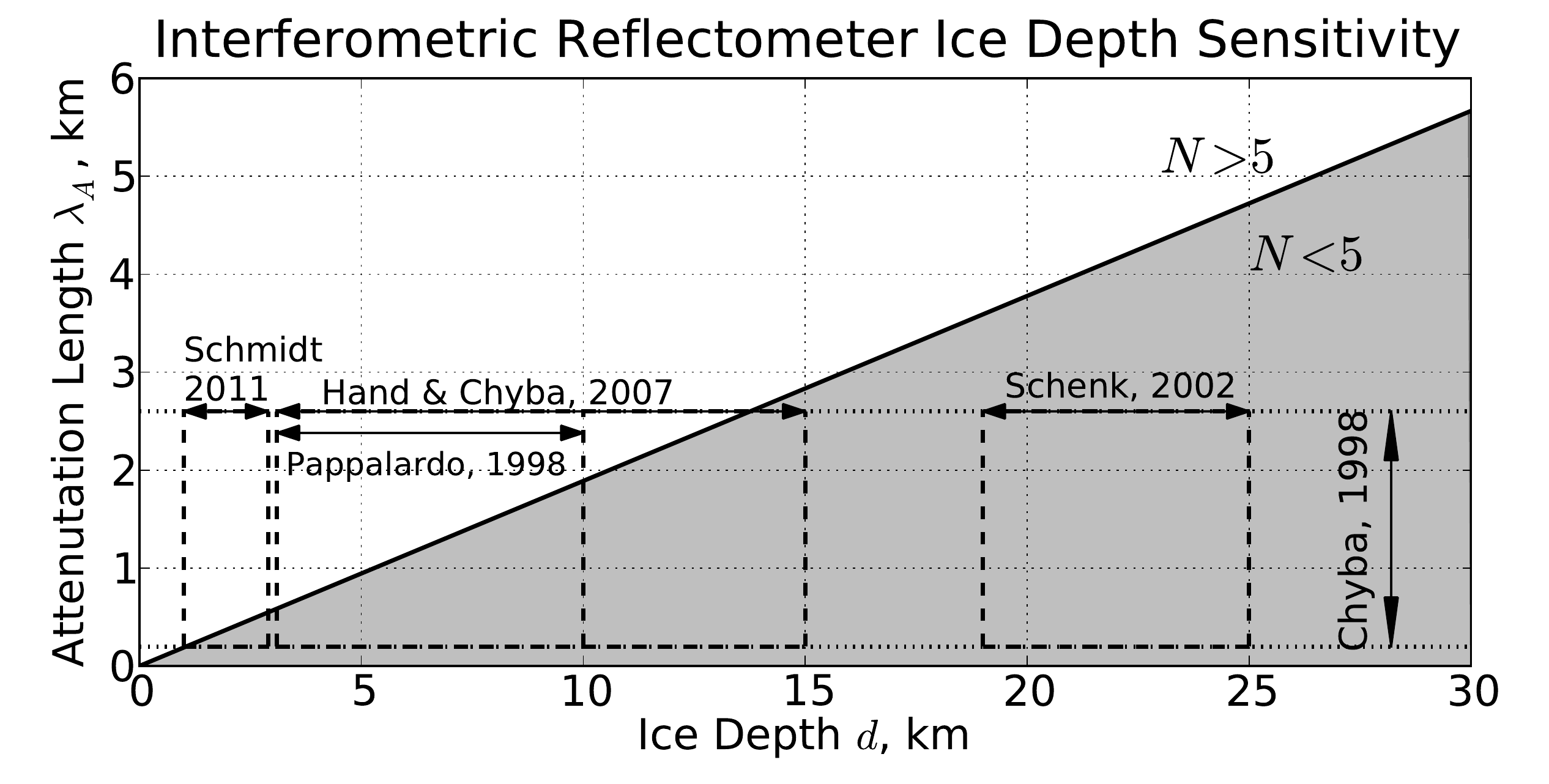}
\caption{Ice depth sensitivity for a 100~km orbiter observing at 10~m wavelengths with a 3~MHz bandwidth. The maximum integration time of 0.5~seconds is assumed. Parameters below the shaded region result in interferometric peaks below $N=5$ times the noise. If the ice attenuation length and depth are in the unshaded region, then observations of a subsurface ocean with high statistical significance can be made. The attenuation length estimates of Chyba, 1998 are shown by the two-sided vertical arrow. The boxes show the ice depth predictions of several studies labeled by their reference.} 
\label{fig:sens_ana} % caption for the whole figure
\end{figure*}

Figure~\ref{fig:sens_ana} shows a plot of the $N=5$ limit (Equation~\ref{eqn:sensitivity}) with integration time $\Delta t_{max}=0.7$~s as a function of attenuation length $\lambda_A$ and ice depth $d$. The gray shaded region shows the parameter space that is not detectable with a $N=5$ sensitivity. The unshaded region is the portion of the parameter space that is detectable above with $N>5$ sensitivity. As discussed in Section 3.2, the modeled attenuation lengths range between 0.2 and 2.6~km. With the estimates made here this means that the deepest subsurface reflector that could be detected with a signal that is $N=5$ times the noise is $<12$~km. This technique is capable of unambiguously detecting the shallow ice crust thicknesses $d\lesssim 3-10$~km  discussed by Pappalardo et al., 1998 and Schmidt, 2011 but not the depths of $d>$19-25~km discussed by Schenk, 2002. Note that this is the sensitivity to a single measurement. It is likely that a sequence of measurements for a slowly varying ocean depth 
would require a smaller value of $N$ for an unambiguous observation.%However, it is possible that observing at lower frequencies may significantly improve ice depth sensitivity required.

\subsection{Ice Feature Resolution}

In this section we estimate the limiting resolution effects of the interferometric reflectometer. These ultimately depend on the temporal resolution of the system. In practice, the delta-function $\delta(\tau)$ in Equations~\ref{eqn:autocorrelation_terms}, \ref{eqn:cc_atm_ice}, \ref{eqn:cc_ice_ocn}, and \ref{eqn:cc_atm_ice_ice_ocn} will be approximated by a sinc function
\begin{equation}
\delta(\tau)\to \Delta f \frac{\sin\left(\pi\Delta f \tau\right)}{\pi\Delta f \tau}.
\end{equation} 
The time-domain width of the sinc function is given by $\Delta \tau = (2\Delta f)^{-1}$. The depth resolution depends on the time resolution via $\Delta d=c\Delta \tau$. Thus, for a 3~MHz bandwidth, the depth resolution is $\sim50$~meters. Reducing the bandwidth to 1~MHz results in $\sim150$~meter depth resolution.

A potentially stronger limiting factor on the temporal resolution is the spatial extent of the decametric radio source. If the source subtends a large solid angle in the sky, then the reflected delays will become smeared resulting in wider correlation peaks.

\begin{figure*}[h!]
\centering
\includegraphics[width=0.6\linewidth]{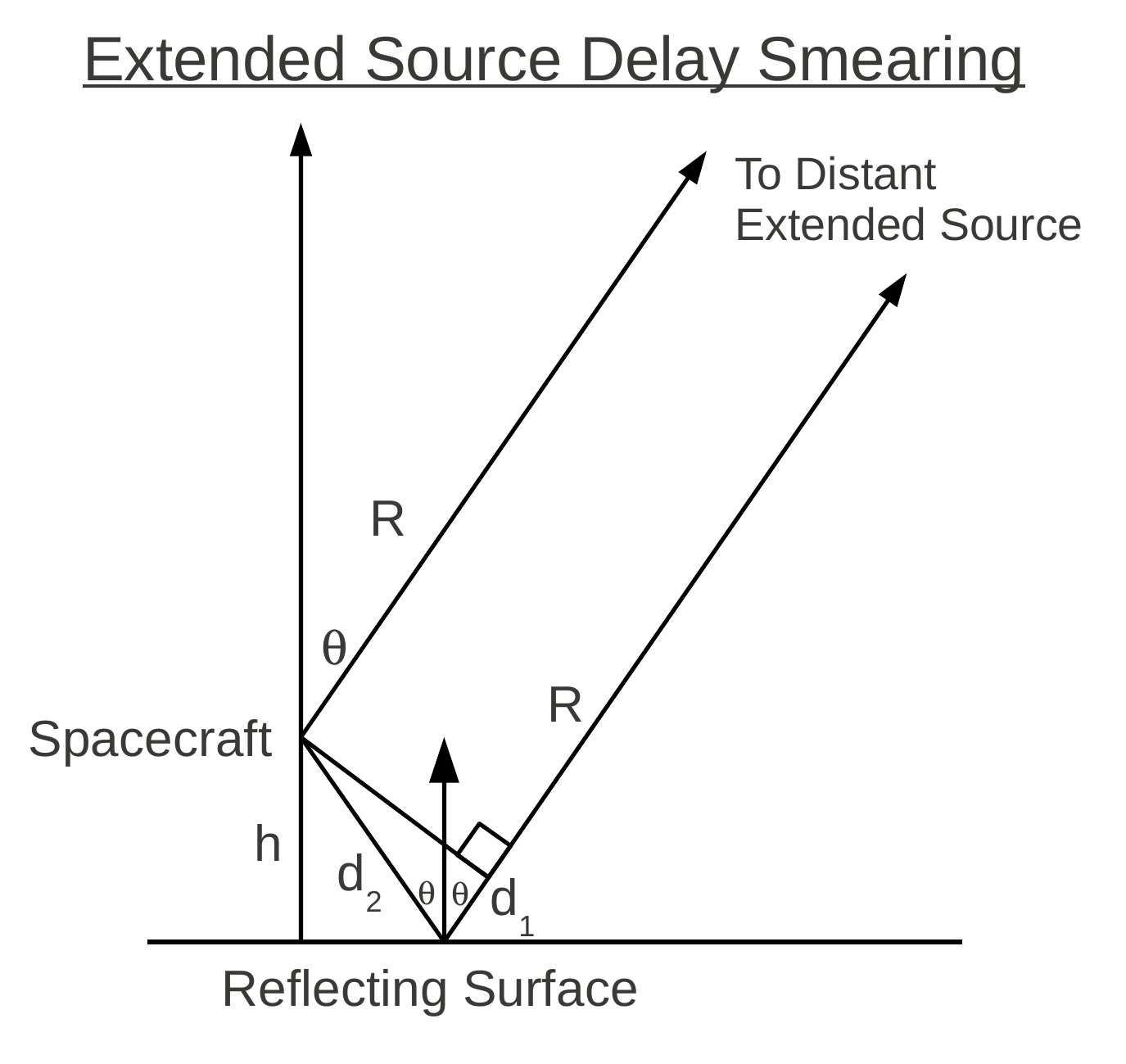}
\caption{The geometry of delay smearing due to an extended source. The delay difference between the direct emission and its reflection is given by the path difference $d_1+d_2$. The emitter lies at $\theta=0^{\circ}$ with an extended source structure such that its radiation also arrives at larger angles $\theta$. The delay smearing is given by Equation~\ref{eqn:delay_smearing}. } 
\label{fig:spatial_smearing} % caption for the whole figure
\end{figure*}

If the source is extended over an angle $\theta$, then the reflections for different portions of the source will arrive over range of delays, potentially interfering with each other and smearing the signal. We can estimate the effect of source spatial extent on time resolution. Let us treat the case where the source is centered on the sub-Jovian point, as shown in Figure~\ref{fig:spatial_smearing}, and extends over an angle $\theta$ small enough where ground curvature effects are negligible. If part of the radiation comes at incident angle $\theta$ with respect to the sub-Jovian axis then the distance of propagation difference between the direct observation and its reflection is given by the sum of distances $d_2=h/\cos\theta$ and $d_1=d_2\cos2\theta$ resulting in a delay
\begin{equation}
\tau=\frac{h}{c}\left(\frac{1+\cos2\theta}{\cos\theta}\right).
\end{equation}

For a spacecraft orbiting over the sub-Jovian point, the delay smear will vary between $\theta=0$ and $\theta=\theta_{max}$ corresponding to the solid angle subtended by the source in the sky, $\Omega_{src}=2\pi(1-\cos\theta_{max})$. The delay smear due to the edges of the disk on the sky, $\Delta\tau=2(\tau(0)-\tau(\theta_{max}))$, is given by
\begin{equation}
\Delta\tau=\frac{2h}{c}\left(2-\frac{1+\cos2\theta_{max}}{\cos\theta_{max}}\right).
\label{eqn:delay_smearing}
\end{equation}

At Europa, the surface of Jupiter has a solid angle with corresponding $\theta_{max}=5.9^{\circ}$. This would give $\Delta\tau=7$~$\mu$s for a spacecraft altitude $h=100$~km. This situation would limit the depth resolution of a a reflective surface to $\sim$2~km. However, the VLBI observations of the decametric emission by Dulk 1970, Carr et al., 1970, and Lynch et al., 1976 have bound the emitting region to a spatial extent $<400$~km, which is much smaller than Jupiter's radius of 69,911~km. At Europa, this corresponds to $\theta_{max}<.034^{\circ}$, which contributes a delay smear of $\Delta\tau<0.2$~ns equivalent to a depth uncertainty of .07~meters. This is well below the depth resolution limit from a $\Delta f\sim3$~MHz bandwidth measurement and is a negligible effect on the ice depth resolution.

The lateral resolution is dominated by the integration time and the motion of the spacecraft. At spacecraft altitudes of 100~km and decametric wavelengths result in Fresnel zones of order 1~km and the maximum integration time is limited to this region. Thus, an orbiting interferometric reflectometer would have  kilometer scale pixels on the surface.

% \begin{figure*}[h!]
% \centering
% \includegraphics[width=0.7\linewidth]{./figures/detection_potential_3.pdf}
% \caption{} 
% \label{fig:correlation} % caption for the whole figure
% \end{figure*}
% 

% \begin{figure*}[h!]
% \centering
% \includegraphics[width=0.7\linewidth]{./figures/refl_corr_20db.pdf}
% \caption{} 
% \label{fig:correlation} % caption for the whole figure
% \end{figure*}

\section{Comparison to Radar Measurements}
In this section we make a first order comparison between an interferometric reflectometer and an ice penetrating radar instrument. We parametrize the performance according to the treatment of Cecconi et al., 2012, with the notation adapted to the one used in this paper
\begin{equation}
P_{Rx}=P_{Tx}\frac{\lambda^2G^2\tau_p\left(1-\rho_{atm-ice}\right)^2 \rho_{ice-ocn}L_{sys}}{\left(4\pi\right)^2\left(2\left(h+d\right)\right)^2} \ \Delta f  \ e^{-2d/\lambda_A},
\end{equation}
where $P_{Rx}$ is the power received from the subsurface ocean reflection, $P_{Tx}$ is the transmitter power (20~Watts), $\lambda$ is the wavelength of the radiation (10~meters), $G$ is the gain (1.5), $\tau_p$ is the transmitter pulse width (150~$\mu$s), $\left(1-\rho_{atm-ice}\right)^2$ is 0.85, $\rho_{ice-ocn}$ is 0.46 for a liquid water ocean, $L_{sys}$ are the system losses ($\sim$0.5), $h$ is the spacecraft altitude (100~km) and $d$ is the depth of a subsurface ocean. With these values $P_{Rx}\sim(2.1\times10^{-8} \ \mbox{Watts/MHz})\Delta f \ e^{-2d/\lambda_A} $
% $P_{Rx}\sim2.7\times10^{-6}e^{-2d/\lambda_A} $~Watts. 

The noise levels for the anti-Jovian side, where Jupiter is occulted by the icy moon, are dominated by the galactic background. The noise power is $P_{N,AJ}=k_{B}T_{A,Gal}\Delta f$. For $T_{Gal}\sim6\times10^4$~Kelvin, at 30~MHz we have 
$P_{N,AJ}=(8.3\times10^{-13} \ \mbox{ Watts/MHz})\Delta f$. The signal to noise ratio in the anti-Jovian side is $SNR_{AJ}=P_{Rx}/P_{N,AJ}\sim2.5\times10^4 e^{-2d/\lambda_A}$. 
% $P_{N,AJ}=2.4\times10^{-12}$~ Watts. The signal to noise ratio in the anti-Jovian side is $SNR_{AJ}=P_{Rx}/P_{N,AJ}\sim1.1\times10^6 e^{-2d/\lambda_A}$. 

On the sub-Jovian side, where Jupiter is in the field of view of the radar antenna, the noise temperature increases to values as high as $T_{A,J}\sim10^{9}$~Kelvin. This results in a noise power of $P_{N,SJ}=(1.4\times10^{-8} \ \mbox{ Watts/MHz}) \Delta f$. The signal to noise ratio in the sub-Jovian side is $SNR_{SJ}=P_{Rx}/P_{N,SJ}\sim 1.5 e^{-2d/\lambda_A}$. This is, however, a worst case estimate since Cecconi et al., 2012 claims that observations can be planned around time when the peak decametric emission from Jupiter has low activity. In any case, the optimal performance of radar will be in the anti-Jovian side.

\begin{figure*}[h!]
\centering
\includegraphics[width=0.9\linewidth]{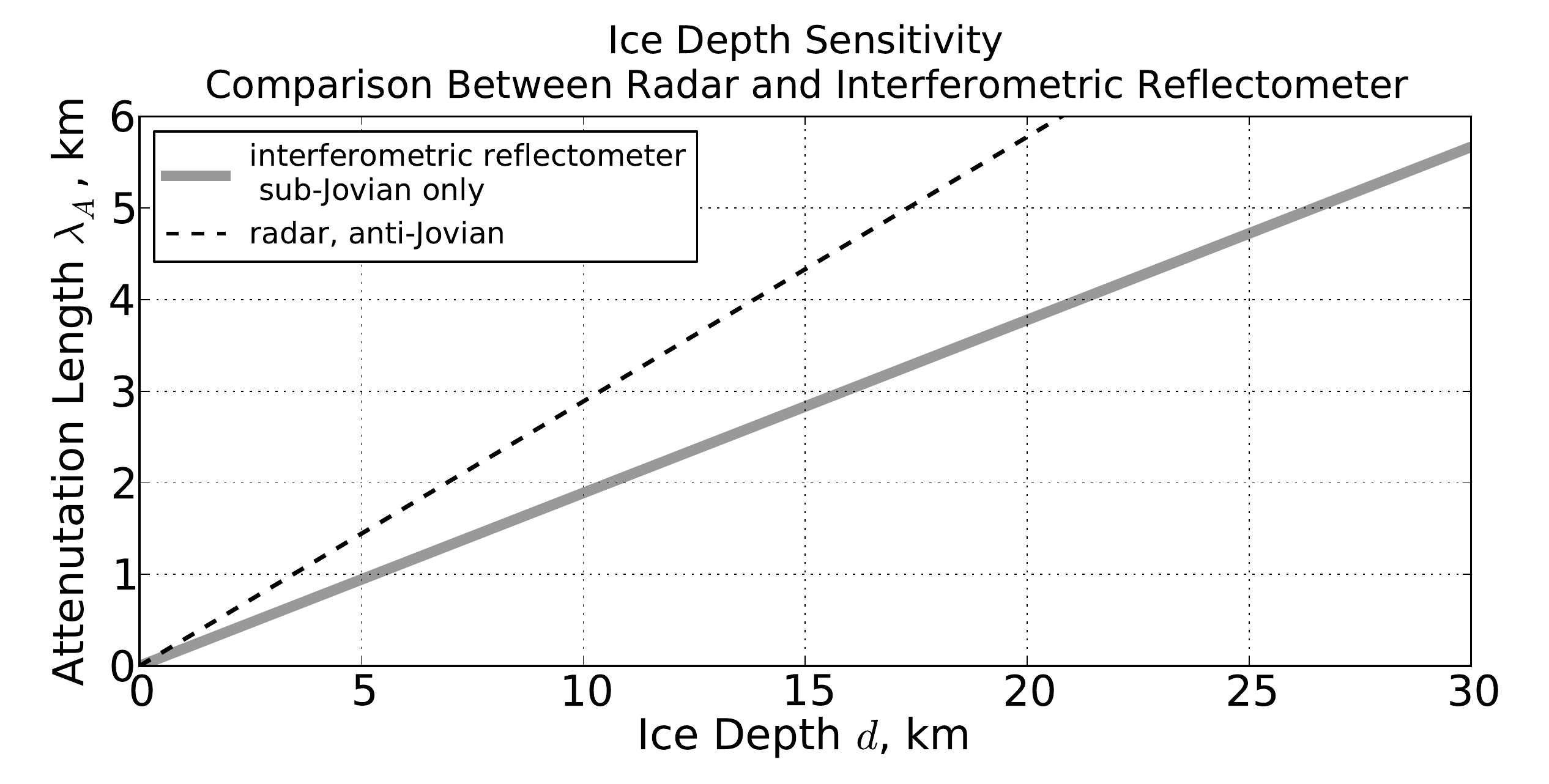}
\caption{Comparison of subsurface ocean sensitivity between interferometric reflectometry in the sub-Jovian side with ice penetrating radar in the anti-Jovian side. Each line is for a $N=5$ detection sensitivity. Ice penetrating radar in the anti-Jovian side, where Jupiter's decametric radiation is occulted, displays comparable sensitivity to interferometric reflectometer in the sub-Jovian side, where IPR is most strongly affected by decametric emission. The techniques are highly complementary for maximizing coverage of an icy moon, especially if the moon is tidally locked. To see where current models and constraints of ice depth and attenuation length lie on this plot see Figure~\ref{fig:sens_ana}.} 
\label{fig:comparison} % caption for the whole figure
\end{figure*}

We can compare the sensitivities of these radar estimates with the interferometric reflectometer. The requirement for observing a signal peak that is $N=5$ times above the noise for a radar measurement is given by
\begin{equation}
\frac{d}{\lambda_A}<\frac{1}{2}\log\left(\frac{SNR}{25}\right)
\end{equation}
where we take the natural logarithm. The sensitivity contours for the sub-Jovian and anti-Jovian sides are shown in Figure~\ref{fig:comparison}. All the contours shown are for a $N=5$ detection, as in Figure~\ref{fig:sens_ana}. 

The results indicate that an interferometric reflectometer could provide a significant increase in sensitivity in the sub-Jovian side. Equally optimistic assumptions have been made in both cases, where surface roughness and clutter effects have not been included. 
% It is possible that radar may be able to operate with a stronger transmitter and perhaps improve the 0.5 system losses included by Cecconi et al., 2012. But note that the sensitivity contour is a logarithmic function of these parameters. 
The techniques are highly complementary for maximizing coverage of an icy moon, especially if the moon is tidally locked, as is the case for Europa.

\section{Outlook and Conclusions}

We have provided the physical basis for a passive interferometric reflectometer that takes advantage of Jupiter's strong decametric emission to search for subsurface oceans in Jovian icy moons. We have shown that the absorptive properties of the ice could allow for enough of this signal to be reflected back to a spacecraft for passive observation of a subsurface ocean. The unambiguous detection of a subsurface ocean could be obtained with a relatively simple system consisting of a dipole antenna, a digitizer, and a correlator.

The interferometric reflectometer concept could be used as a complementary system to an ice penetrating radar instrument by adding a passive device sharing the radar antenna. Radar provides its best measurements in the anti-Jovian side while the interferometric reflectometer works best in the sub-Jovian side. This is also a very low power system that could run in the background while other instruments are performing their measurements.

There have been a number of simplifying assumptions in this work that will have to be studied in more depth. In particular, we have assumed that Jupiter's decametric radio emission behaves as white noise. Partial coherence in the bursts could improve the correlation compared to a white noise model. The autocorrelation behavior of the Jovian decametric emission can be constrained in a future study using data from low frequency arrays such as LOFAR and the LWA.

Other features of Jovian decametric emission need to be included in the interferometric reflectometer measurement model. More detailed simulations using specific orbits and modeled behavior of the different components of the decametric emission and their spatio-temporal characteristics will be required. The Cyclotron Maser Instability model of decametric radiation (Treumann, 2006) claims that the decametric radiation originates in the poles of Jupiter and the emission propagates as a cone with wide opening angle but narrow width. The emission comes both from the north and south poles, which could potentially extend the region where interferometric reflectometry is applicable. Since some of the stronger emissions are sporadic, it also may be advisable to include a power meter that triggers the correlator when there is a sudden boost in the decametric flux density.

Another possibility is that the interferometric reflectometer can work with an electrically short dipole. The dimensions of the antenna used for low frequency measurements can be a limiting factor in the design of an instrument. A resonant dipole antenna for frequencies around 3~MHz implies a dimension comparable to a half wavelength of $\lambda/2=50$~meters, which is impractical for a deep space probe. At a frequency of 30~MHz, a dipole antenna has dimension of $\lambda/2=5$~meters, which is more manageable. A 5~meter dipole operating as an electrically short antenna could extend observations of decametric radiation down to frequencies as low as 3~MHz. This would be particularly advantageous given the wide band over which decametric activity is observed and where the surface roughness and clutter effects are largely reduced. The extension to lower frequencies could provide significant improvements in sensitivity and should be studied in the future.

The spatial, temporal, and spectral structure of the decametric emission needs to be studied in more detail for future instrument development. Many of the details left out in this first study could improve the estimates for this technique and open the way for a passive radio probe for Solar System exploration.

\section*{Acknowledgements}
We would like to thank Mike Janssen, Sam Gulkis, Steve Levin, Chuck Naudet, Charley Dunn, Jim Zumberge, Luis Amaro, and William Smythe at the Jet Propulsion Laboratory for their helpful and encouraging discussions on this idea. 
We would also like to thank Imke de Pater at UC Berkeley for her helpful advice. This research was carried out at the Jet Propulsion Laboratory, California Institute of Technology, under a contract with the National Aeronautics and Space Administration and funded through the Internal Research and Technology Development program. Copyright 2014 California Institute of Technology. Government sponsorship acknowledged.

\appendix

\section{Geometric Optics of Plane Waves Reflected Off a Spherical Surface}

We will use geometric optics to derive an approximate analytical formula for the scattering of plane waves off a spherical surface. Here we will consider the geometric factors only. The losses due to reflection geometry are estimated by taking a plane wave ray bundle with area $A_1$ and calculating the spread of rays after reflection into an area $A_2$. 

\begin{figure*}[h!]
\centering
\includegraphics[width=0.9\linewidth]{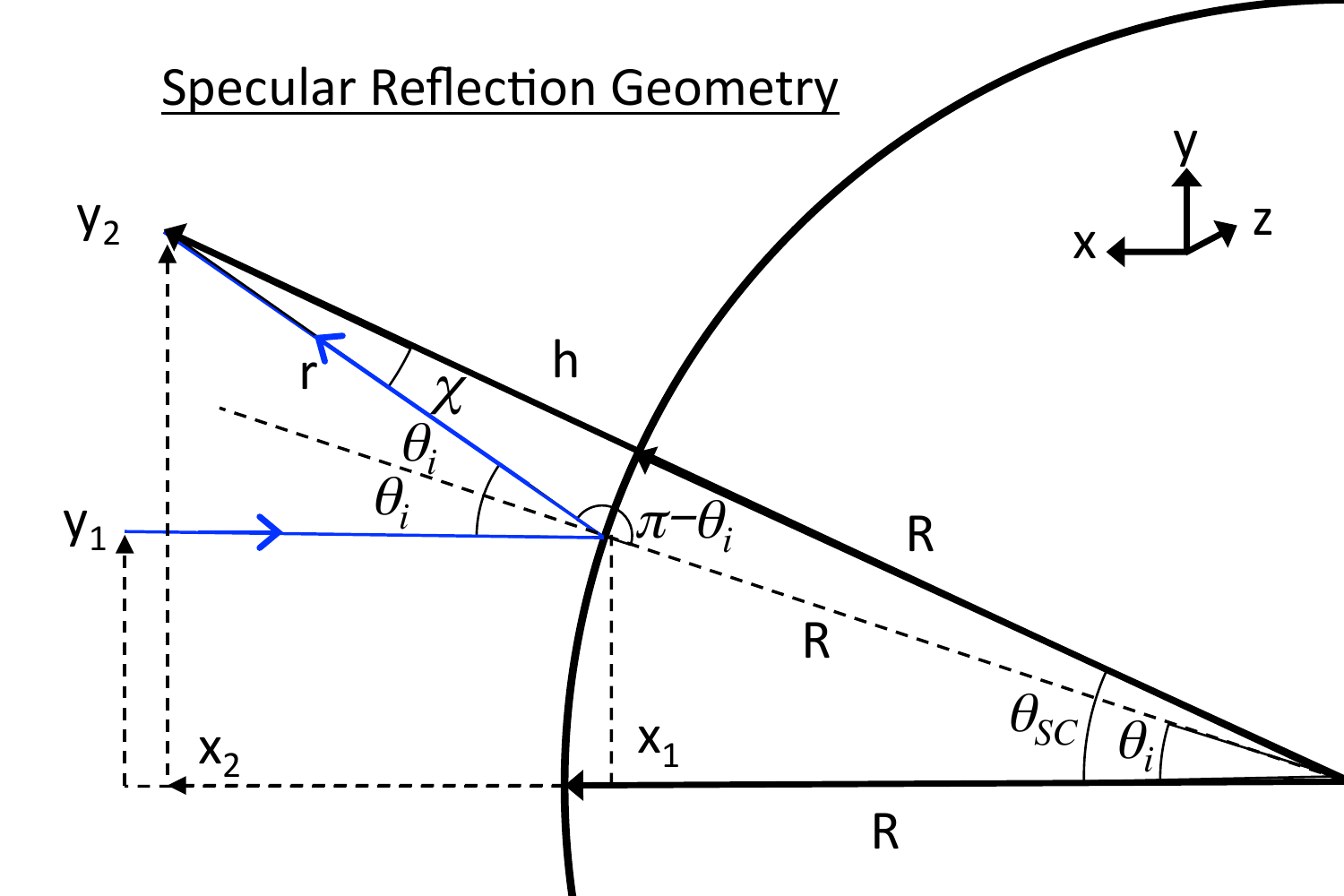}
\caption{
Geometry of specular reflection of plane wave incident on the surface of a sphere of radius $R$ with angle $\theta_i$ as observed from a spacecraft with altitude $h$ above the surface. The spacecraft is located at angle $\theta_{SC}$. The reflection point lies at an angle $\chi$ from the nadir of the spacecraft.
} 
\label{fig:scat_geom} % caption for the whole figure
\end{figure*}

Before we estimate the geometric spread of a reflected plane wave let us derive some basic relations. Figure~\ref{fig:scat_geom} shows the scattering geometry for a plane wave aligned with the horizontal axis incident on a spherical surface of radius $R$ with angle $\theta_i$. The specular reflection off the surface of the sphere propagates the ray to a spacecraft at height $h$ and angle $\theta_{SC}$ with respect to the plane wave axis.

In Cartesian coordinates, the location of the reflection point on the sphere is given by 
\begin{equation}
(x_1,y_1)=(R\cos\theta_i, R\sin\theta_i)
\label{eqn:x1y1}
\end{equation}
while the coordinates of the spacecraft are given by
\begin{equation}
(x_2,y_2)=((R+h)\cos\theta_{SC},(R+h)\sin\theta_{SC}).
\label{eqn:x2y2}
\end{equation}
The spacecraft nadir angle $\chi$ to the reflection point is given by 
\begin{equation}
\sin\chi=\frac{\sin\theta_i}{1+h/R}.
\label{eqn:chi}
\end{equation}
The spacecraft angle $\theta_{SC}$, the nadir angle to reflection $\chi$, and the incidence angle $\theta_i$, are related to each other via
\begin{equation}
\theta_{SC}=2\theta_i-\chi.
\label{eqn:theta_SC}
\end{equation}

\begin{figure*}[h!]
\centering
\includegraphics[width=0.9\linewidth]{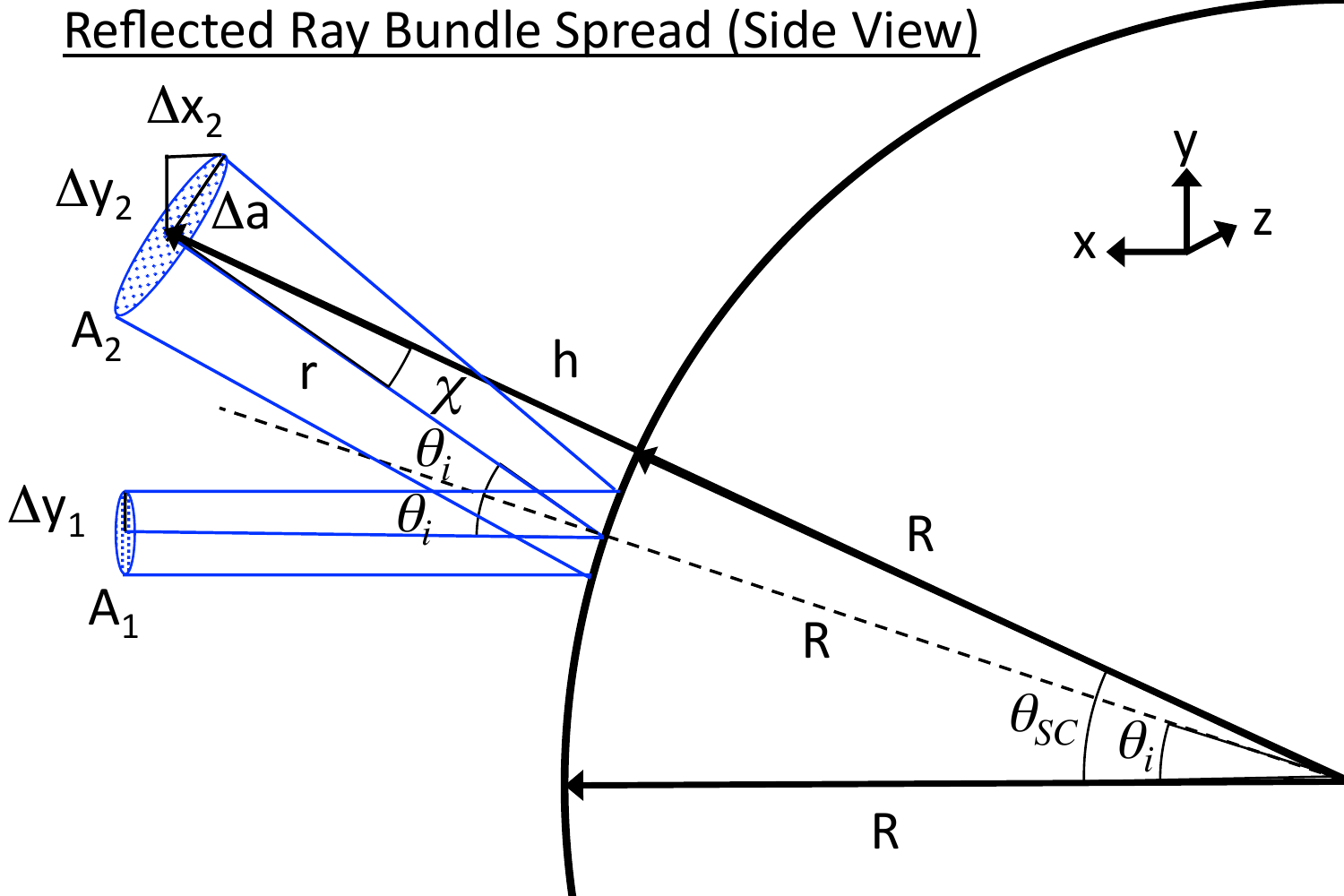}
\caption{
Geometry of a ray bundle with area $A_1$ incident on the surface of a sphere and spread into an area $A_2$. $\Delta y_1$ is circular radius of the incoming bundle. The reflected shape is spread into an approximately elliptical shape with semi-major axis $\Delta a$. The $x,y$ components of $\Delta a$, in the coordinate system shown in this figure, are $\Delta x$ and $\Delta y$. 
} 
\label{fig:scat_geom_2} % caption for the whole figure
\end{figure*}

The geometry of the ray bundle reflection and spreading is shown in Figure~\ref{fig:scat_geom_2}. The incoming bundle has area $A_1=\pi\Delta y_1^2$ which, upon reflection, is spread to an area $A_2$. We estimate $A_2$ by approximating it as an ellipse. In the plane of incidence the ellipse has semi-major axis $\Delta a$, which is given in terms of the projected distances $\Delta x_2, \Delta y_2$ by $\Delta a=\sqrt{\Delta x_2^2 + \Delta y_2^2 }$.

We can obtain $\Delta x_2, \Delta y_2$ via $\Delta x_2=(\partial x_2 / \partial y_1)\Delta y_1$  and $\Delta y_2=(\partial y_2 / \partial y_1)\Delta y_1$. From the geometrical relations in Equations~\ref{eqn:x1y1}~and~\ref{eqn:x2y2}, combined with the chain rule with $\theta_i$ as an intermediate variable, the partial derivatives are given by
\begin{equation}
\frac{\partial x_2}{\partial y_1}=\frac{(1-\cos\theta_i) + (1+2h/R)}{\cos\theta_i}\sin\theta_{SC}
\end{equation}
\begin{equation}
\frac{\partial y_2}{\partial y_1}=-\frac{(1-\cos\theta_i) + (1+2h/R)}{\cos\theta_i}\cos\theta_{SC}
\end{equation}
Together these equations give
\begin{equation}
\Delta a=\frac{(1-\cos\theta_i) + (1+2h/R)}{\cos\theta_i}\Delta y_1
\end{equation}

\begin{figure*}[h!]
\centering
\includegraphics[width=0.9\linewidth]{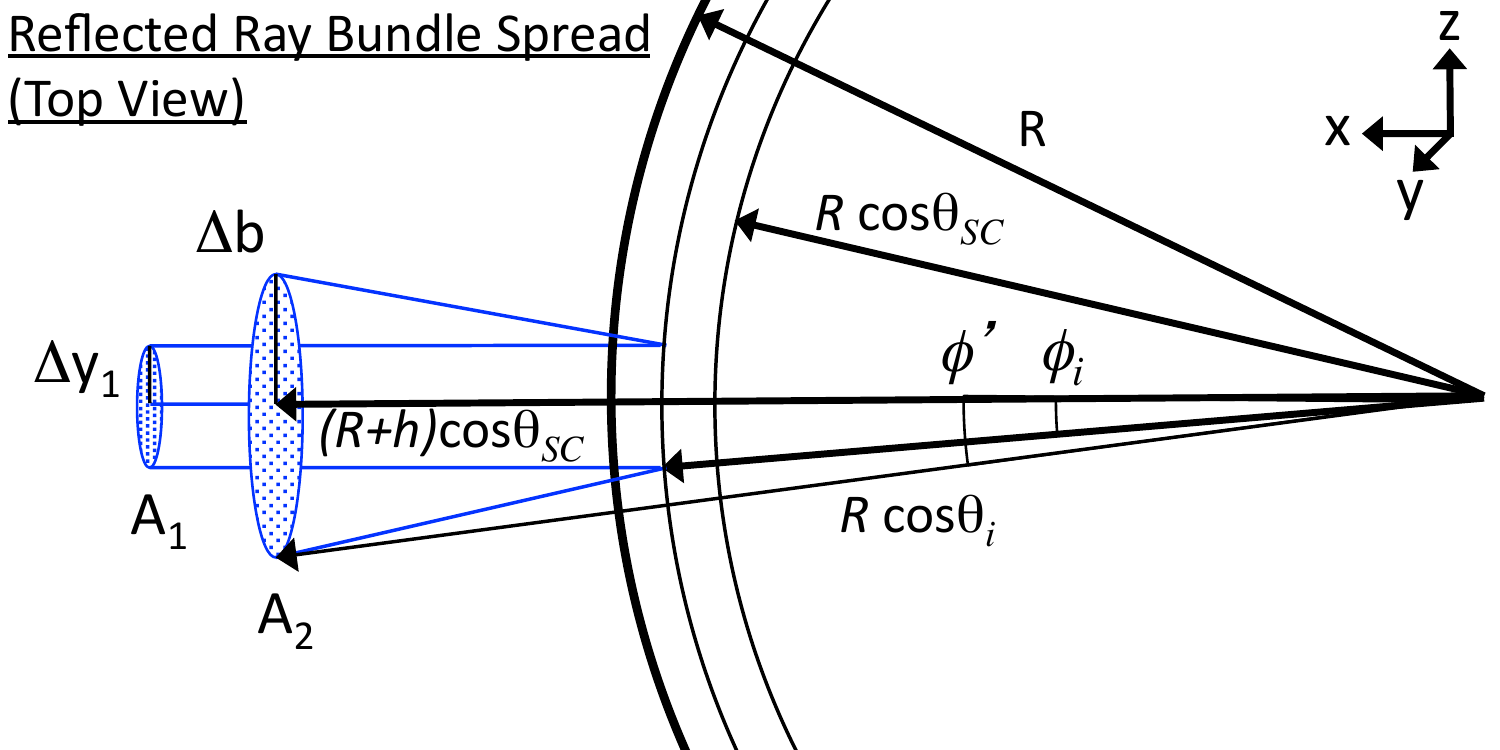}
\caption{
Rotated view of Figure~\ref{fig:scat_geom_3}. In this projection, the area ellipse $A_2$ has semi-minor axis $\Delta b$.
} 
\label{fig:scat_geom_3} % caption for the whole figure
\end{figure*}

The geometry for the calculation of the semi-minor axis $\Delta b$ of the area ellipse $A_2$ is illustrated in Figure~\ref{fig:scat_geom_3}, which is the top view of Figure~\ref{fig:scat_geom_2}. The are three contours on the sphere that correspond to the projected radii of $R$, $R\cos\theta_i$, and $R\cos\theta_{SC}$. From figure Figure~\ref{fig:scat_geom_3} it is readily seen that 
\begin{equation}
\Delta b=(R+h)\cos\theta_{SC}\tan\phi'
\label{eqn:Delta_b}
\end{equation}
where $\phi'$ is the azimuthal angle of the vector pointing from the center of the sphere to the edge of the semi-minor axis of ellipse $A_2$. We also have 
\begin{equation}
\Delta y_1=R\cos\theta_i\tan\phi_i
\label{eqn:delta_y_1}
\end{equation}
where $\phi_i$ is the azimuthal angle of the vector pointing from the center of the sphere to the edge of the area $A_1$ enclosing the incoming ray bundle.
Thus finding $\Delta b$ in terms of $\Delta y_1$ reduces to the problem of finding $\phi'$ in terms of $\phi_i$. 

Finding $\phi'$ in terms of $\phi_i$ requires that we use vectors in three dimensions. Let $\mathbf{\hat{k}}$ be the direction of the incoming ray bundle and $\mathbf{\hat{n}}$ the unit vector normal to the surface of the sphere at the reflection point. Let $\mathbf{\hat{k}}'$ be unit vector in the direction of the reflected ray given by 
\begin{equation}
\mathbf{\hat{k}}'=\mathbf{\hat{k}}-2(\mathbf{\hat{k}}\cdot\mathbf{\hat{n}})\mathbf{\hat{n}}
\end{equation}

We want to find the unit vector $\mathbf{\hat{n}}'$ pointing from the center of the sphere to the spacecraft at altitude $h$. The azimuthal angle $\phi'$ is given by
\begin{equation} 
\tan\phi'=\mathbf{\hat{n}}'\cdot\mathbf{\hat{y}}/\mathbf{\hat{n}}'\cdot\mathbf{\hat{x}}
\label{eqn:tan_phi_prime}
\end{equation}. 
Where the $x$-axis is given by the direction of the incoming plane wave and the spacecraft lies in the $x-z$ plane. We can find $\mathbf{\hat{n}}'$ using the relation
\begin{equation}
(R+h)\mathbf{\hat{n}}' = R\mathbf{\hat{n}} + r\mathbf{\hat{k}}'
\label{eqn:unit_vector_relation}
\end{equation}
where $r$ is the distance between the reflection point and the spacecraft, as shown in Figure~\ref{fig:scat_geom_2}. Squaring the magnitude on both sides of Equation~\ref{eqn:unit_vector_relation} and solving for $r$ gives
\begin{equation}
r=-R\mathbf{\hat{n}}\cdot\mathbf{\hat{k}} + R\sqrt{\left(\mathbf{\hat{n}}\cdot\mathbf{\hat{k}}\right)^2+\left(1+h/R\right)^2-1}
\end{equation}
where we have substituted $\mathbf{\hat{n}}\cdot\mathbf{\hat{k}}'=-\mathbf{\hat{n}}\cdot\mathbf{\hat{k}}$. We can now solve for 
\begin{equation}
\mathbf{\hat{n}}' = \frac{R}{R+h}\mathbf{\hat{n}} + \frac{r}{R+h}\mathbf{\hat{k}}'
\end{equation}
Using $\mathbf{\hat{k}}=(1,0,0)$ and $\mathbf{\hat{n}}=(\cos\theta_i\cos\phi_i,\cos\theta_i\sin\phi_i,\sin\theta_i)$ we have $\mathbf{\hat{n}}\cdot\mathbf{\hat{k}}=\cos\theta_i\cos\phi_i$, which gives
\begin{equation}
r=-R\cos\theta_i\cos\phi_i + R\sqrt{\cos^2\theta_i\cos^2\phi_i+\left(1+h/R\right)^2-1}.
\end{equation}
The reflected vector direction is given by
\begin{equation}
\mathbf{\hat{k}}'=
\left(
\begin{array}{c}
1-2\cos^2\theta_i\cos^2\phi_i \\
-2\cos^2\theta_i\cos\phi_i\sin\phi_i \\
-2\cos\theta_i\sin\theta_i\cos\phi_i
\end{array}
\right).
\end{equation} 
This gives 
\begin{equation}
\mathbf{\hat{n}}'\cdot\mathbf{\hat{x}}=\frac{1}{1+h/R}\left(\cos\theta_i\cos\phi_i+\frac{r}{R}\left(1-2\cos^2\theta_i\cos^2\phi_i\right)\right)
\label{eqn:n_prime_dot_x}
\end{equation}
and
\begin{equation}
\mathbf{\hat{n}}'\cdot\mathbf{\hat{y}}=\frac{1}{1+h/R}\left(\cos\theta_i\sin\phi_i-\frac{r}{R}\left(2\cos^2\theta_i\cos\phi_i\sin\phi_i\right)\right)
\label{eqn:n_prime_dot_y}
\end{equation}
Substituting Equations~\ref{eqn:n_prime_dot_x}~and~\ref{eqn:n_prime_dot_y} into Equation~\ref{eqn:tan_phi_prime} and using a first order small angle approximation for $\phi_i$ near $\phi_i\approx\pi$ and $\phi'$ near $\phi'\approx\pi$ we have the relation
\begin{equation}
\phi'\approx
\frac{
\cos\theta_i+(r/R)\left(2\cos^2\theta_i\right)
}
{
\cos\theta_i-(r/R)\left(1-2\cos^2\theta_i\right)
}
\phi_i
\end{equation}
Substituting Equation~\ref{eqn:delta_y_1} into this expression and the result into Equation~\ref{eqn:Delta_b} gives 
\begin{equation}
\Delta b = \left(1+h/R\right)\left(\frac{\cos\theta_{SC}}{\cos\theta_i}\right)\frac{
\cos\theta_i+(r/R)\left(2\cos^2\theta_i\right)
}
{
\cos\theta_i-(r/R)\left(1-2\cos^2\theta_i\right)
}\Delta y_1
\end{equation}

The ratio of the area of the reflected bundle $A_2=\pi\Delta a \Delta b$ to the area $A_1=\pi\Delta y_1^2$ of the incoming bundle is
\begin{equation}
\begin{split}
\frac{A_2}{A_1} &= \left[\frac{(1-\cos\theta_i) + (1+2h/R)}{\cos\theta_i}\right] \\
&\times
\left[
\left(1+h/R\right)\left(\frac{\cos\theta_{SC}}{\cos\theta_i}\right)\frac{
\cos\theta_i+(r/R)\left(2\cos^2\theta_i\right)
}
{
\cos\theta_i-(r/R)\left(1-2\cos^2\theta_i\right)
}
\right].\\
\end{split}
\label{eqn:A2A1}
\end{equation}

The geometric losses for a flat plate collector are given by the ratio $A_1/A_2$. However, since we are dealing with antennas, we want the equivalent geometric losses for an isotropic collector. To correct for this, we divide the ratio $A_1/A_2$ by $\cos\chi$, which gives the angle of the incoming rays with respect to the normal of the flat plate with area $A_2$ (see Figure~\ref{fig:scat_geom_2}). The geometric factor $g(R,h,\theta_i)$ for an isotropic antenna is given by 
\begin{equation}
g(R,h,\theta_i)=\frac{1}{\cos\chi}\frac{A_1}{A_2},
\label{eqn:geom_fac_calculation}
\end{equation}
where $\chi$ is given in terms of the incident angle $\theta_i$ by Equation~\ref{eqn:chi} and $A_1/A_2$ is given in terms of $\theta_i$ by the inverse of Equation~\ref{eqn:A2A1}, with $\theta_{SC}$ given in terms of $\theta_i$ by Equation~\ref{eqn:theta_SC}.

As a check on the analytical approximation, we have simulated the reflection with ZEMAX (\texttt{http://www.radiantzemax.com}). The results are shown in Table~\ref{tbl:zemax_results}. The simulations and analytical results for $A_1/A_2$ agree to within a few percent in the region of interest to interferometric reflectometry.
\begin{table}
\centering
\begin{tabular}{|c|c|c|c|c|c|c|}
\hline
$h/R$ & 
\begin{tabular}[x]{@{}c@{}}$\theta_i$,\\degrees\end{tabular} & 
\begin{tabular}[x]{@{}c@{}}$\chi$,\\degrees\end{tabular} & 
$r/R$ & \begin{tabular}[x]{@{}c@{}}geometric factor \\(analytical) \end{tabular}  &   \begin{tabular}[x]{@{}c@{}}$A_1/A_2$\\(analytical)\end{tabular} & 
\begin{tabular}[x]{@{}c@{}}$A_1/A_2$\\(ZEMAX)\end{tabular}   \\
\hline
0.2 & 0   & 0    & 0.2  & 0.510    & 0.510  & 0.512\\
1   & 0   & 0    & 1    & 0.111    &  0.111 & 0.110\\
4   & 0   & 0    & 4    & 0.0123   & 0.0123 & 0.0123\\
0.2 & 7.5 & 6.2  & 0.2  & 0.506    & 0.503  & 0.493\\
0.2 & 15  & 12.5 & 0.21 & 0.494    & 0.482  & 0.482\\
0.2 & 30  & 24.6 & 0.23 & 0.447    & 0.406  & 0.392\\
\hline
\end{tabular}
\caption{ZEMAX simulation results for estimating the geometric factor. The columns are the ratio of the height to the radius of the sphere $h/R$, the angle of incidence on the sphere $\theta$, the spacecraft nadir angle of the reflection point $\chi$, the ratio of the distance from the spacecraft to the reflection point to the radius of the sphere $r/R$, the geometric factor, the ray bundle area spread ratio calculated analytically, and the area ratio calculated with ZEMAX. The analytical estimate and ZEMAX for $A_1/A_2$ agree to within a few percent in the worst case.}
\label{tbl:zemax_results}
\end{table}

Note that for $\theta_i=0^{\circ}$, the geometric factor is 
\begin{equation}
g(R,h,\theta_i=0^{\circ})=\frac{(R/2)^2}{(R/2+h)^2}.
\end{equation} 
In the limit of large $h$ this tends to $h^{-2}$, which is the expected geometric attenuation for a far-field radiator. For small $h$, as considered in this study, the ratio of areas behaves as $1-4h/R$, which is a significantly more slowly varying function than $h^{-2}$.

\section{Time Delays of Surface and Subsurface Reflections}

\begin{figure*}[h!]
\centering
\includegraphics[width=0.5\linewidth]{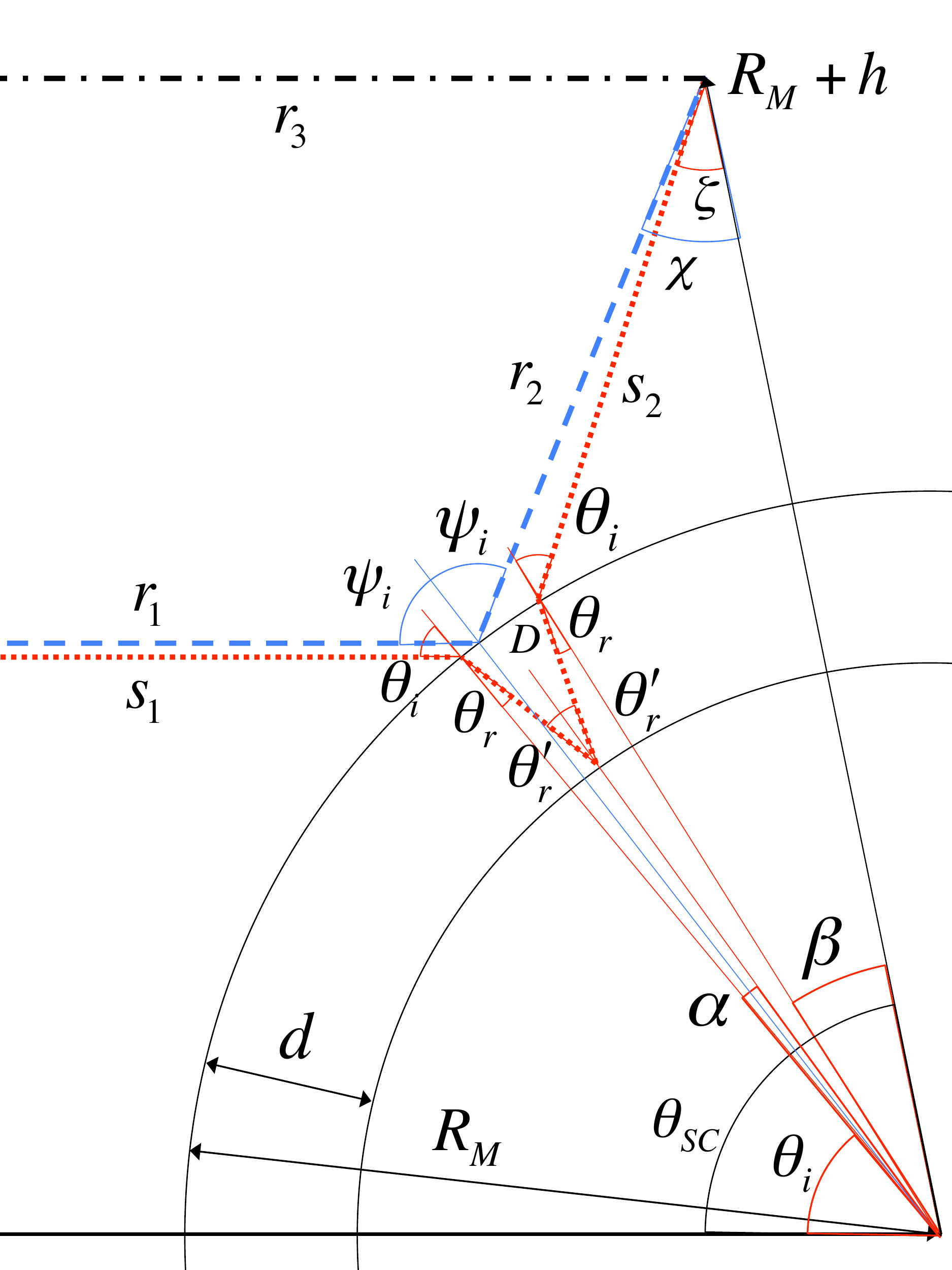}
\caption{
Geometry for calculating the delays of plane wave signals arriving from Jupiter, reflecting off a spherical surface and subsurface layer at depth $d$, and observed at a spacecraft with altitude $h$. The blue dashed line traces a surface reflection with incident angle $\psi_i$ while the red dotted line traces a subsurface reflection with incidence angle $\theta_i$. Both arrive at the spacecraft located at angle $\theta_{SC}$. The subsurface reflection is refracted with angle $\theta_r$, given by Snell's law, and reflects off the subsurface layer with angle $\theta'_r$. The ray propagates a distance $D$ from the surface to the reflection point. The total distance of propagation in the ice is $2D$.
} 
\label{fig:reflections_geom} % caption for the whole figure
\end{figure*}

In this appendix we calculate the expected delays of the surface and subsurface reflected light rays coming from Jupiter. Figure~\ref{fig:reflections_geom} shows the geometry of the surface and subsurface reflected rays as observed from a spacecraft at altitude $h$ and orbit angle $\theta_{SC}$. 

The surface reflected light ray (shown is a dashed blue line) has incident angle $\psi_i$. The delay between the direct Jovian emission $\tau_{atm-ice}$, modeled as a plane wave, and its surface reflection is given by the path length difference
\begin{equation}
\tau_{atm-ice}=(r_{1}+r_{2}-r_{3})/c
\label{eqn:tau_ice_atm}
\end{equation}
where $c$ is the speed of light in vacuum, $r_1$ is the distance from Jupiter to the reflection point, $r_2$ is the distance between the reflection point and the spacecraft, and $r_3$ is the distance from Jupiter to the spacecraft. Let $D_J$ denote the distance from the center of the icy moon of radius $R_M$ to Jupiter. The distances in Equation~\ref{eqn:tau_ice_atm} are given by
\begin{equation}
r_1=D_J-R_M\cos\psi_i
\label{eqn:r1}
\end{equation}
\begin{equation}
r_2=\sqrt{(R_M+h)^2+R_M^2-2R_M(R_M+h)\cos(\theta_{SC}-\psi_i)}
\label{eqn:r2}
\end{equation}
\begin{equation}
r_3=D_J-(R_M+h)\cos\theta_{SC}
\label{eqn:r3}
\end{equation}

The spacecraft angle $\theta_{SC}$ is given in terms of the incident angle $\psi_i$ according to
\begin{equation}
\theta_{SC}=2\psi_i-\arcsin\left(\frac{\sin\psi_i}{1+h/R_M}\right)
\end{equation}
The delay $\tau_{atm-ice}$ as a function of $\theta_{SC}$ is obtained by substituting Equations~\ref{eqn:r1},~\ref{eqn:r2}~and~\ref{eqn:r3} into Equation~\ref{eqn:tau_ice_atm}. Note that $D_J$ vanishes from the result.

For the subsurface reflection (shown in dotted red lines) the delay calculation has to account for the propagation in ice. As shown in Figure~\ref{fig:reflections_geom} the subsurface reflected ray observed at spacecraft orbit angle $\theta_{SC}$ has incident angle $\theta_i$. The light ray refracts into the ice according to Snell's law with angle
\begin{equation}
\theta_r=\arcsin\left(\frac{1}{n_{ice}}\sin\theta_i\right),
\end{equation}
where $n_{ice}$ is the index of refraction of ice. The ray propagates a distance $D$ from the transmission point to the subsurface layer reflecting with angle $\theta'_r$ given by 
\begin{equation}
\theta'_r=\arcsin\left(\frac{\sin\theta_r}{1-d/R_M}\right),
\label{eqn:theta_r_prime}
\end{equation}
where $d$ is the ice depth. By symmetry, the exit reflection, refraction, and transmission angles are the same as the entry angles. The spacecraft orbit angle $\theta_{SC}$ is given by
\begin{equation}
\theta_{SC}=2\theta_i+2(\theta'_r-\theta_r)-\arcsin\left(\frac{\sin\theta_i}{1+h/R_M}\right)
\label{eqn:theta_SC_subsurface}
\end{equation}

The delay between the direct Jovian emission and its subsurface reflection is given by the path length difference
\begin{equation}
\tau_{ice-ocn}=\left(s_1+2n_{ice}D+s_2-r_3\right)/c
\label{eqn:tau_ice_ocn}
\end{equation}
where 
\begin{equation}
s_1=D_J-R_M\cos\theta_i,
\label{eqn:s1}
\end{equation}
\begin{equation}
D=\sqrt{R_M^2+(R_M-d)^2-2R_M(R_M-d)\cos(\theta'_r-\theta_r)},
\label{eqn:D}
\end{equation}
\begin{equation}
s_2=\sqrt{(R_M+h)^2+R_M^2-2R_M(R_M+h)\cos\left(\theta_i-\arcsin\left(\frac{\sin\theta_i}{1+h/R_M}\right)\right)},
\label{eqn:s2}
\end{equation}
and $r_3$ is given by Equation~\ref{eqn:r3}. Substituting these equations into Equation~\ref{eqn:tau_ice_ocn} gives $\tau_{ice-ocn}$ as a function of spacecraft angle $\theta_{SC}$.

\end{document}